\documentclass[useAMS,usenatbib]{mn2e}

\usepackage{graphicx}
\usepackage{times}
\usepackage{natbib}

\newcommand{\apjs}{ApJS}

\newcommand{\apjl}{ApJ Letters}

\newcommand{\aap}{A\&A}

\newcommand{\enzo}{\small ENZO}

\begin{document}
 
\title[Complexity and Information in Cosmic Structures]{On the complexity and the information content of cosmic structures}

\author[F. Vazza]{F. Vazza$^{1,2}$
\thanks{%
 E-mail: franco.vazza@hs.uni-hamburg.de}\\
 %EndAName
$^{1}$ Hamburger Sternwarte, Gojenbergsweg 112, 20535 Hamburg, Germany
$^{2}$ INAF, Istituto di Radioastronomia di Bologna, via Gobetti 101, 41029 Bologna, Italy}

\date{Accepted ???. Received ???; in original form ???}
\maketitle

\begin{abstract}
The emergence of cosmic structure is commonly considered one of the most complex phenomena in Nature. 
However, this {\it complexity} has never been defined nor measured in a quantitative and objective way.
In this work we propose a method to measure the information content of cosmic structure and to quantify
the complexity that emerges from it, based on Information Theory. 
The emergence of complex evolutionary patterns is studied with a statistical symbolic analysis of the datastream produced
by state-of-the-art cosmological simulations of forming galaxy clusters.  This powerful approach 
allows us to measure how many bits of information are necessary to predict the evolution of energy fields in a statistical way, and it offers
a simple way to quantify when, where and how the cosmic gas behaves in complex ways. 
 The most complex behaviors are found in the peripheral regions of galaxy clusters, where supersonic flows drive shocks and large energy fluctuations over a few tens of million years. Describing the evolution of magnetic energy requires at least a twice as large amount of bits than for the other energy fields. When radiative cooling and feedback from galaxy formation are considered, the cosmic gas is overall found to double its 
degree of complexity.  In the future, Cosmic Information Theory can significantly increase our understanding of the emergence of cosmic structure as it represents an innovative framework to design and analyze complex simulations of the Universe in a simple, yet powerful way.

\end{abstract}

\label{firstpage}

\section{Introduction}
\label{sec:intro}

How did the large-scale structure of the Universe come into shape? 
Decades of study have suggest that this complex structure has emerged from a
hierarchy of interconnected processes, where several mechanisms (e.g. the expansion of the space time, gravity, hydrodynamics, radiative and chemical gas processes, etc) have coupled in a non linear way, regulating the flow of energy across scales and leading to the formation of cosmic voids, filaments, galaxies and galaxy clusters \citep[e.g.][]{1993ppc..book.....P, 1985ApJS...57..241E,KA99.3,2005Natur.435..629S,2014Natur.509..177V}. This is arguably one of {\it the most complex} problem in physics (if not the most complex). 
However, there has been to date little attempt to define or measure what {\it complexity} in the formation of cosmic structures really means. 

This is not just a merely speculative question: according to Information Theory \citep[e.g.][for a review]{prokopenko2009information} any physical system - including the Universe itself - can be regarded as an information-processing device, which continuously computes its own evolution. In this view, the set of physical laws relevant to evolve the system is analogous to the software (i.e. the set of rules) used to advance the computation. Understanding where the software behaves in a complex way has the potential to give us a deeper insight on the emergence of self-organization in the Universe around us. \\

Information Theory is a very interdisciplinary and steadily growing research field, whose origin is commonly dated to the seminal work by \citet{1949IEEEP..37...10S} and \citet[][]{1949mtc..book.....S} on signal processing in communications. 
We further refer the reader to \citet{feldman1997bii}, \citet{adami}, \citet{2004PhRvL..93n9902S} and 
\citet{prokopenko2009information} for a few reviews on the topic.
In several other field of physics, there have been valuable attempts to define and study complexity, based on Information Theory, including climate data analysis \citep[e.g.][]{hoffman2011data}, cellular automata \citep[][]{1984PhyD...10....1W}, limnology \citep[][]{2013arXiv1304.1842F}, epidemiology \citep[][]{escidoc:2219209}, and many more. 
Applications of Information Theory to astrophysics concerns the reconstruction of sparse signals \citep[e.g.][]{ens09,ens11,ens13}, cosmology \citep[e.g.][]{2004PhRvL..92n1302H,2012PhRvD..86h3539L}, extragalactic surveys \citep[e.g.][]{2013MNRAS.430.3376P,2015MNRAS.454.2647P} and compact stars  \citep[][]{deAvellar20121085}.
Crucial to these attempts is the consideration that a physical phenomenon can be treated as an information processing device, and its evolution can  be studied through "word" (i.e. symbolic) statistical analysis.

To our knowledge, Information Theory has never been applied before to study the formation of cosmic structures, and in this work we explore for the first time methods to quantify and describe the emergence of complexity in the Universe.
In particular, we first focus on the emergence of the largest self-gravitating structures of the Universe, i.e. galaxy clusters, by mean
of state-of-the-art cosmological simulations.
Indeed, the deep connection between information processing and physical evolution is easy explored by means of numerical simulations, because 
the two aspects are made equivalent by construction. 
This allows us to measure complexity in the simulated data (in $\rm bits$) and to relate it with the underlying gas-dynamical processes that are captured by the simulation.\\
%To best highlight the role of different processes (e.g. non-radiative processes vs radiative cooling and galaxy feedback) we monitor the 
%growth of complexity in two high-resolution resimulations of the same object. \\

The paper is organized as follows: in Section \ref{methods} we give a schematic overview of the algorithms from Information Theory that are relevant to our modeling of complexity and information in cosmological simulations. In Section \ref{zeld} we discuss the preliminary application of these methods to a simple 1-dimensional simulation of structure formation, while in Sec.\ref{cosmo} we analyze the complexity of 3-dimensional simulations of galaxy clusters. To best highlight the role of different processes (e.g. non-radiative processes vs radiative cooling and galaxy feedback) in making the intracluster medium complex,  we compare two resimulations of the same object in Sec.\ref{subsec:rad}. In Sec.\ref{eff_icm} we contrast the macroscopic and the microscopic view of complexity in the intracluster medium,  while in Sec.\ref{conclusion} we give our conclusions.
In the Appendix we present additional tests on the algorithms used in the main paper.

\section{Methods}
\label{methods}

\subsection{Information and complexity}
\label{info}
We first give a basic overview of the methods used in this work to measure the information content of numerical simulations.
For further details, we refer the interested reader to the excellent review by \citet[][]{prokopenko2009information}. 

\subsubsection{Shannon's information entropy}

Information theory states that the information content related to the outcome a probabilistic process, $x$, with probability $P$, can be defined as $log_2[1/P]=-log_2[P]$ \citep{1949mtc..book.....S}, whose unit of measure is the {\it bit}.
This measure is known as {\it information entropy} and it measures the amount of freedom of choice (or the degree of randomness) contained in the process. 
Intuitively, this states that a process with many possible outcomes has high entropy  and  this measure is suitable to quantify ''how much choice'' is involved in the selection of the event and/or of how uncertain we are of the outcome. 
%Manifestly,  the information content of a dataset does not account per-se for the {\it complexity} of the dataset itself. 
%This become obvious considering that the same given amount of bit of a dataset can encode very informative simulation data or complete noise statistics. 
Therefore, according to this interpretation, {\it the complexity of a physical system equals to the amount of information needed to describe its evolution.}

\subsubsection{The algorithmic complexity}
The minimal information needed to perfectly describe the system is measured by the {\it algorithmic complexity} \citep[e.g.][]{kolm,1995chao.dyn..9014C}. In computer simulations, this is basically set by the disk memory necessary to store every single digit produced by the simulation itself or, alternatively, the entire source code and its initial conditions. 
Qualitatively speaking, the inherent complexity of structure formation is made evident by the fact that 
long and complex algorithms are necessary to produce to a realistic simulation of how large-scale structures evolve 
{\footnote{We give some examples from a few widely used codes in cosmology: the  Lagrangian smoothed-particle-hydrodynamical code {\small GADGET-2} (http://www.mpa-garching.mpg.de/gadget/)  has a compressed size of  $\sim 200 \rm kb$  in its basic version, while for a similar amount of physical routines the Eulerian adaptive mesh refinement method {\small RAMSES}  (http://www.ics.uzh.ch/\~teyssier/ramses/) has a compressed size of $\sim 600 \rm kb$. The latest version of the adaptive mesh refinement code {\small ENZO} (https://code.google.com/p/enzo/), which we will use in this work, has a compressed size of $\sim 2.1 \rm Mb$, including  the numerical routines for magneto-hydrodynamics and radiative transfer.}}.

However, this representation of complexity poses some practical problem, which are best explained by thinking to it as to a compression problem{\footnote{http://www.ics.uci.edu/\~dan/pubs/DataCompression.html}}. A simple periodic object requires very little algorithmic complexity as it can be very significantly compressed, i.e. the necessary source code can be extremely short (e.g. the generator of a sine function). Conversely, a total random sequence of digits has no 
 internal structure and cannot be described but by storing every single element, i.e. the only possible loss-less compression of this the data is the data itself. 

This definition of complexity does not fully reflect our intuition of what is really complex in Nature. Indeed, even a manifestly more complex sequence of elements like the sequence of pressure fluctuations in a fluid, the ensemble of orbits of a planetary systems, or even a novel or a 5-voices fugue by J. S. Bach can be more compressed than a random sequence of digits. For this reason, other alternative approaches to characterize
complexity have been developed.

\subsubsection{Statistical complexity}

From a more physical viewpoint, what is relevant is to 
 quantify  how much  information is necessary to {\it statistically} describe the evolution of a system. This is given by the {\it statistical complexity} \citep[e.g.][]{adami}, which measures how likely it is that a system does many different things at a given time. The statistical complexity also quantifies the similarity between different realizations of the same process. A purely random process is not statistically complex as it always repeats the same patterns in a statistical way. At the same time, in many cases it is reasonable to expect that two different numerical realization of statistically similar initial conditions are characterized by a similar level of complexity, even if their final outputs are punctually different. This is often the case encountered in cosmological simulations, which produce different sample of objects (e.g. galaxies) with 
statistically similar properties if they start from statistically similar sets of initial conditions.
 
The statistical complexity is usually measured by partitioning the system into discrete levels ($E_i$, with $1 \leq i \leq N_{\rm bin}$, $N_{\rm bin}$ being the total number of levels in the partition) and by calculating the conditional probability distribution that elements at a given level at an epoch $t$, $E_i(t)$, transition to another level at a following epoch $t +\Delta t$, $E_j(t+\Delta t)$. The $N_{\rm bin} \times N_{\rm bin}$ matrix of all possible transition at each epoch is directly measured in the datastream, and hence the transition probability distribution {\bf $P[E_j(t+\Delta t)|E_i(t)]$}.  

Each spatial element of the system is therefore regarded as a processing unit, responsible for the production of a stream of $L$ symbols (where $L$ is the total number of epochs/timesteps) drawn from a ``vocabulary'' of $N_{\rm bin}$ words (i.e. energy levels). At any given timestep, each spatial element in the system (identified by its 3D position $xyz$)
is characterized by the transition probabilty associated to its evolution, $P_{xyz} \equiv P_{xyz}[E_j(t+\Delta t)|E_i(t)]$.
 It is worth stressing that such matrix of transitions is directly derived from the datastream at every timestep, without requiring any prior knowledge of the underlying dynamics.
The Shannon entropy associated to the probability of transition for each $xyz$ element gives its statistical complexity (in bits): 

\begin{equation}
C_{\mu,xyz} = - P_{xyz} \log_2 P_{xyz}
\label{eq:complex}
\end{equation}

\noindent while the total of this over the domain gives the total statistical
complexity of the system 

\begin{equation}
C_{\mu}=\sum_{xyz} C_{\mu,xyz}.
\label{eq:complex_tot}
\end{equation}

In this view, every element of the system acts as an information-processing unit producing a datastream of "symbols", and the conditional probability of transitions between symbols quantifies how much complex is the evolution that the underlying system is computing.
The statistical complexity can also be seen as the typical information needed to produce a sequence of symbols statistically similar 
to the original sequence of symbol of the system being studied. 
In the following, we will always refer our estimate of statistical complexity to the time lag that separes a given epoch to the previous output of the simulation, i.e. $\Delta t$ will be always the time separation between the two last timesteps considered.
  
\subsubsection{Block entropy and source entropy rate}
\label{subsubsec:block_entropy}

The entire sequence of $L$ symbols/states in long time-series carries information that is accessible to an observer. To this end, one needs to extract from the datastream the probability distribution of all sequence of symbols verified in the system, $W^L \equiv W(X^L)$, where $X^L$ denotes the collection of all sequences of $L$ symbols actually occurring in the datastream \citep[e.g.][]{Larson20111592}. Therefore, the probability of occurrence of a specific sequence of $L$ symbols for any $xyz$ element in the computing domain is drawn from the entire $X^L$ collection of all sequences with the same $L$-length, which occur in the datastream.
The complexity associated of a given $L$-sequence of symbols for each computing element $xyz$ is then given by the {\it block entropy}, $H_{xyz}(L)$ :

\begin{equation}
H_{xyz}(L) = - W_{xyz}^L \log_2 W_{xyz}^L 
%\sum_{W^L \in W^L} p(w^L) \log p(w^L),
\label{eq:block_entropy}
\end{equation}

\noindent measured in bits, with $W_{xyz}^L \in W^L$. 
The total block entropy of the system is therefore obtained by summing across the domain:
\begin{equation}
 H(L)= \sum_{xyz} H_{xyz} (L).
\label{eq:block_entropy_tot}
\end{equation}

In these specific case of cosmological simulations, the sequence of symbols is made by coarse graining the  energy levels in the resolution elements (i.e. cells) of the simulation as a function of time. 
The block entropy is a monotonically increasing function of the symbol length  \citep[e.g.][]{feldman1997bii,Crutchfield03}, and the
the increase of the block entropy with $L$ is measured by the {\it entropy gain}:

\begin{equation}
h_\mu(L)=H(L)-H(L-1), 
\label{gain}
\end{equation}

\noindent measured in bits per symbol. 

This metric converges to the same estimate of  $H(L)/L$ in the limit of large $L$, giving the {\it source entropy rate}:
\begin{equation}
h_\mu=\lim_{L\rightarrow \infty} h_\mu(L)=\lim_{L\rightarrow \infty}  H(L)/L.
\end{equation}

The entropy gain is a good identifier of intrinsic randomness in a sequence of symbols as it estimates the information-carrying capacity in the $L$-blocks that is not actually random, but is instead due to correlations. In the practical applications explored in this work, $L$ is not arbitrarirly large but is limited to the maximum amount of timesteps, which is of order $\sim 300-400$ (Sec.\ref{zeld}-\ref{cosmo}), and therefore our estimate of $h_\mu$ from the previous
equation is an approximation.

\subsubsection{Excess entropy and efficiency of prediction}
\label{subsubsec:eff}

As suggested by \citet{2003Chaos..13...25C}, the total apparent memory of structure in a source of $L$ symbols can be quantified through the
{\it excess entropy}: 

\begin{equation}
E= \sum_{L=1\rightarrow \infty} [h_\mu (L)-h_\mu].
\end{equation}

The excess entropy measures the amount of information at a specific value of $L$ that is ``explained away'' by measuring
correlations over larger and larger blocks, i.e. it measures the intrinsic redundancy in the sources of symbols.
For practical applications, $E$ can be simplified into a finite partial-sum  for a length $L$:

\begin{equation}
E(L)=H(L) - L \cdot h_\mu (L),
\end{equation}

\noindent as we will in this work.

Systems with a large dynamical range can be studied on different scales. Most astrophysical objects have this property, and they
can be characterized on several different scales, by averaging or coarse graining their internal states on different scales of interests.
This often leads to the natural question whether there is a level which is best to derive 
a workable model capable of quantitative predictions for a system under analysis.
\citet[][]{2004PhRvL..93n9902S} suggested that  this is addressed by the {\it efficiency of prediction}: 

\begin{equation}
e=\frac{E}{C_\mu},
\end{equation}

\noindent simply computed  as the ratio between the excess entropy and the statistical complexity. 
The scales at which the ratio $e$ is maximum defines the scale at which making predictions of the future evolution of the system is more efficient. Indeed, while the excess entropy $E$ gives the amount of information that can be used to predict the future evolution of a system, given its past, the complexity $C_\mu$ gives the amount of information needed to statistically reproduce a process.
Therefore, the ratio $e=E/C_\mu$  is a good estimate of ``how much can be predicted'' compared to ``how much difficult it is to predict'' \citep[][]{prokopenko2009information}.
The scale at which $e$ is maximum (by construction, $e \leq 1$) defines the spatial or the temporal scale at which the datastream originated from the system displays the maximum {\it emergence} of coherent structures.

%\begin{figure}
%\includegraphics[width=0.49\textwidth]{images/temp_dens.eps}
%\caption{Evolution of gas density, temperature and entropy for the Zeldovich collapse test.} 
% \label{fig:zeld1}
%\end{figure}

\begin{figure}
\includegraphics[width=0.459\textwidth]{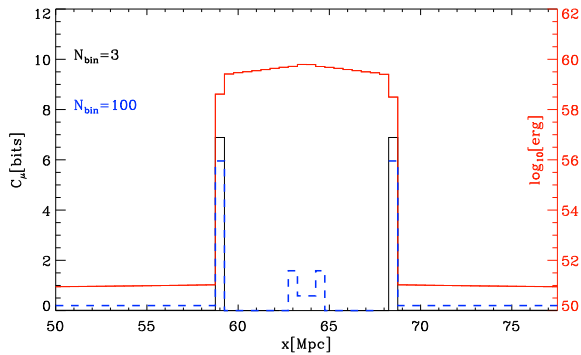}
\caption{Statistical complexity at $z=0$ for the Zeldovich collapse test (measured using $N_{\rm bin}=3$, in black, or $N_{\rm bin}=100$, in blue) and gas energy (red) at $z=0$}
 \label{fig:zeld_complex}
\end{figure}

\begin{figure}
\includegraphics[width=0.459\textwidth]{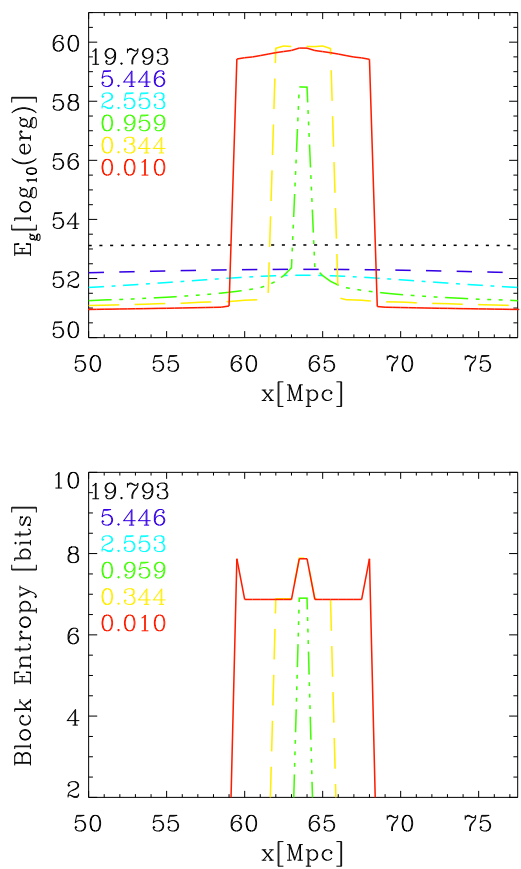}
\caption{Evolution of the gas energy and of the block entropy in the Zeldovich collapse test.} 
 \label{fig:zeld2}
\end{figure}

\section{Results}

\subsection{One dimensional structure formation:  the Zeldovich pancake}
\label{zeld}

%We present here the results of the application of the above quantitative tools to measure the information content and complexity of simulated large-scale structures. 
We first test the above applications  Information Theory to the study of structure formation in one dimension, 
using the Zeldovich collapse test (also known as the "Zeldovich pancake" \citealt[][]{1972A&A....20..189S}). 
Despite its simplicity, this 1-dimensional problem gives a robust representation of the basic physics ruling the growth of cosmic structures originating from a uniform smooth density background. 
In this problem, a uniform density cold gas is initialized with a small converging velocity profile, which later induces the formation of a self-gravitating  center of mass where gas matter continues to accrete.  The first evolutionary stage is adiabatic and the  temperature smoothly increases towards the center due to compression, while in a later stage the accretion velocity becomes supersonic and strong accretion shocks are generated around the central "pancake", leading to the efficient thermalization of infall kinetic energy. No dark matter and no dark energy are included in this simple model, yet the final evolutionary stage of the pancake gives a reasonable representation of the hot and over-dense gas (hereafter intracluster medium, ICM) of a galaxy cluster. We used a $256$ cells domain with a comoving size of $128$  Mpc and initial redshift of $z=30$, for a flat Universe only made of baryons $\Omega_{\rm b}=1$, simulated with the same code used in the following Sections, the cosmological 3-D grid code {\enzo} \citep[][]{enzo14}. 

The collapse of the pancake is followed by supersonic accretion around $z \sim 1$ and at this epoch strong ($\mathcal{M} \gg 10$) accretion shocks are
formed marking the transition between the free-falling gas and the thermalised pancake, at the critical cosmological density.
The final temperature of the pancake is $\sim 10^7-10^8 \rm K$, with an innermost density peak which has a smaller
temperature than the average to maintain pressure equilibrium with the surrounding gas. 

Here and in the following cosmological simulations, we focus on the evolution of the simulated energy fields to measure the growth of complexity as outlined above (Sec.\ref{methods}).
Our choice is to partition the internal state of the simulated gas into energy levels: the comoving kinetic energy ($E_K$), the comoving thermal energy ($E_T$) and the comoving magnetic energy ($E_B$). The dynamical energy range of these fields is so large across the cosmic volume, i.e. more than $\sim 10$ orders of magnitude, that we must adopt a coarse binning in the logarithmic energy space.  In this first test, we just discuss the evolution of gas energy for clarity, while in the full 3-dimensional case (Sec.\ref{cosmo}) we also consider the kinetic and the magnetic energy fields. Figure \ref{fig:zeld_complex} gives the final energy configuration of the pancake at $z=0$ (red line), while the upper panel of Fig.\ref{fig:zeld2} shows the evolution of $E_T$ at different redshifts. 

In detail, the gas energy is defined as $E_T=3k_{\rm B}T \rho dV/(2 \mu m_p)$,  where $dV$ is the volume of the cell and $\mu$ is the mean molecular weight (here $\mu=0.6$). $m_p$ and $k_{\rm B}$ are the proton mass and the Boltzmann constant, respectively. We discretise the gas energy into $N_{\rm bin}$ logarithmic energy bins, and tested the extreme cases of $N_{\rm bin}=3$ and $=100$ (blue and black lines in Fig.\ref{fig:zeld_complex}).\\

The statistical complexity, $C_\mu$ (Eq. \ref{eq:complex}), is measured by reconstructing the 
 matrix of all measured transitions between two simulated timesteps ($\Delta t \approx 30$ Myr). Based on this datastream, we compute for each cell $x$ the transition probability between $E_T (x,t)$ and $E_T(x,t+\Delta t)$, defined as  $P(E_T(x,t+\Delta t)|E_T(x,t))$.
The statistical complexity $C_\mu$ for the entire system is computed
using Eq.\ref{eq:complex}-Eq.\ref{eq:complex_tot}.\\

 Figure \ref{fig:zeld_complex} gives the statistical complexity for each cell in the simulated Zeldovich pancake at $z=0$.  Different choices of the energy binning are expected to give the same information on the strongest energy transitions in the system, but to lead to an increased detail in the more subtle energy transitions. On the other hand, smaller energy bins also make the algorithm more prone to enhance spurious numerical fluctuations of energy levels, which can also be connected to the (de)refining of the grid in adaptive mesh calculation (see following Section).

We find that the highest statistical complexity in the pancake is at accretion shocks, where $E_T$ transitions in a few timesteps from the extremes of the energy distribution. The results for $N_{\rm bin}=3$ or $=100$ show that this behavior is very robust against the adopted
energy binning. This suggest that the surface where gas matter is being thermalized through strong accretion shocks truly represents {\it the most complex location} in this cosmic structure.   On the other hand, only with a very fine coarse graining of the energy variable, a complex behavior is also measured in the core of the pancake, due to the fluctuations
 of gas thermal energy following  gas compression. With the finest binning of $N_{\rm bin}=100$,  even the rarefied outer wings of accreting gas display a small complexity, as an effect of rarefaction waves in the gas. It shall be noticed that at least a part of the complexity here actually measures fluctuations of numerical origin, which are expected in region undergoing strong rarefaction.\\
 
 The block entropy $H(L)$  (Sec.\ref{subsubsec:block_entropy}) is built by 
 analyzing the full sequence of $N_{\rm step}=322$ time steps of the simulation.
 For each cell in the 1-D domain we constructed the sequence of $N_{\rm step}$ symbols for the logarithmically binned values of  of $E_T$
 (here using $N_{\rm bin}=3$ to reduce the number of symbols). From this, we draw the complete distribution of 
 all possible sequences of symbols with length $L$ occurred in the recorded datastream, $X^L$.
 This way we can compute the a posteriori probability of each sequence of symbols occurred in the  datastream, {\bf $W(X^L)$}, and from this deriving
  the block entropy for each cell as a function of epoch (Eq. \ref{eq:block_entropy}).
The evolution of the block entropy for each cell, $H_{xyz}(L)$, is given in the bottom panel of  Figure \ref{fig:zeld2}.
  
Only the late evolution of the pancake, $z \leq 1$, shows a complex enough sequence of symbols to be detected by our $N_{\rm bin}=3$ choice. 
The evolution of outside of the pancake shows a simple behavior over time, while the maxima of block entropy are again found at accretion shocks.
Approximately  $\approx 7.5 ~\rm bits$ of information are necessary to predict the average evolution of these regions.
Due to its time-based formulation, the block entropy can better highlight the complexity that was involved in the formation of the pancake core, where strong shocks were first formed at $z \sim 1$. In comparison, the statistical complexity view cannot access this information as it only access transitions over a fixed time scale. 
We give in the Appendix the results of different choices of $N_{\rm bin}$, which show that all most important features highlighted in the block entropy analysis are also recovered if a different binning of energies is used.

%In the next Section we will see that most of the above basic trends are also recovered in the full 3-dimensional cosmological case, and therefore
%can offer a qualitative understanding of the general trends of more realistic scenarios.

%\begin{figure*}
%\includegraphics[width=0.99\textwidth]{images/Nat1.ps}
%\caption{Slice through the centre of a $\sim 1.5 M_{\odot}$ cluster at $z=0$, showing the magnetic energy (left), the thermal energy (centre) and the kinetic energy (right).}
 %\label{fig:map0}
%\end{figure*}

\begin{figure*}
\includegraphics[width=0.99\textwidth]{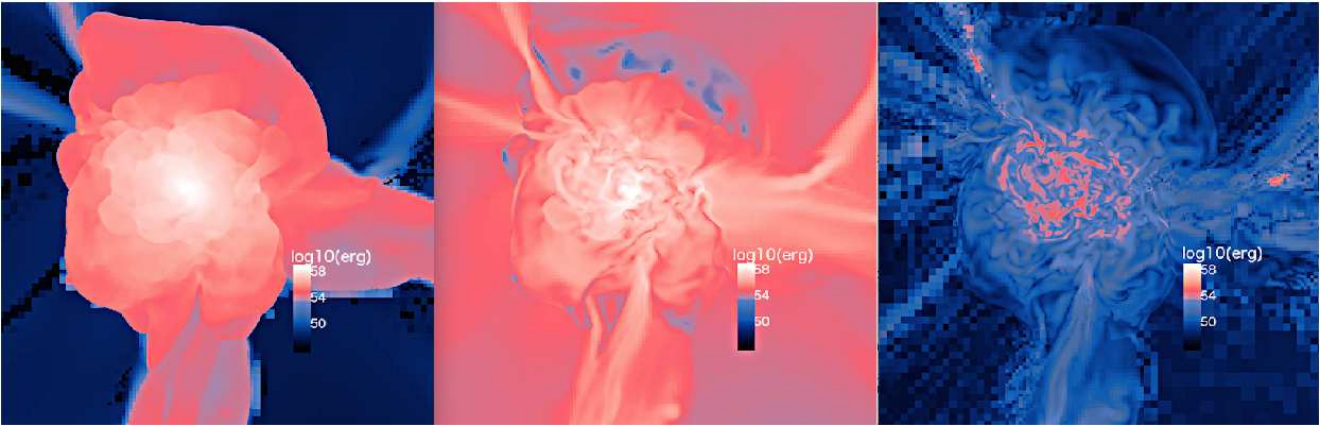}
\caption{Slice through the center of a $\sim 1.1 \cdot 10^{15} M_{\odot}$ cluster at $z=0$, showing the kinetic energy (left), the thermal energy (center) and the magnetic energy (right). Each panel is $15 \times 15 \rm Mpc^2$ across. }
 \label{fig:map0}
\end{figure*}

\begin{figure*}
\includegraphics[width=0.33\textwidth]{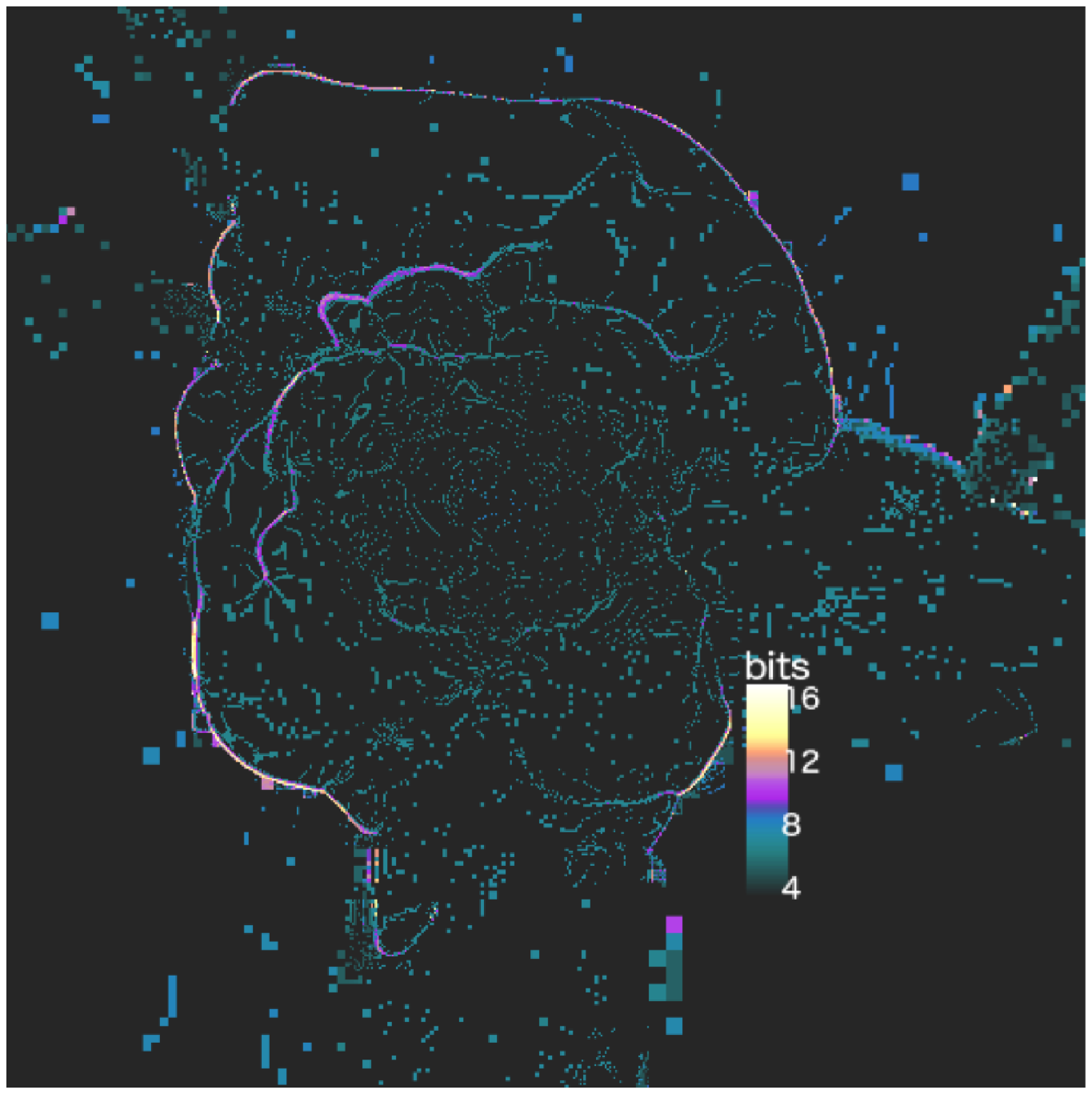}
\includegraphics[width=0.33\textwidth]{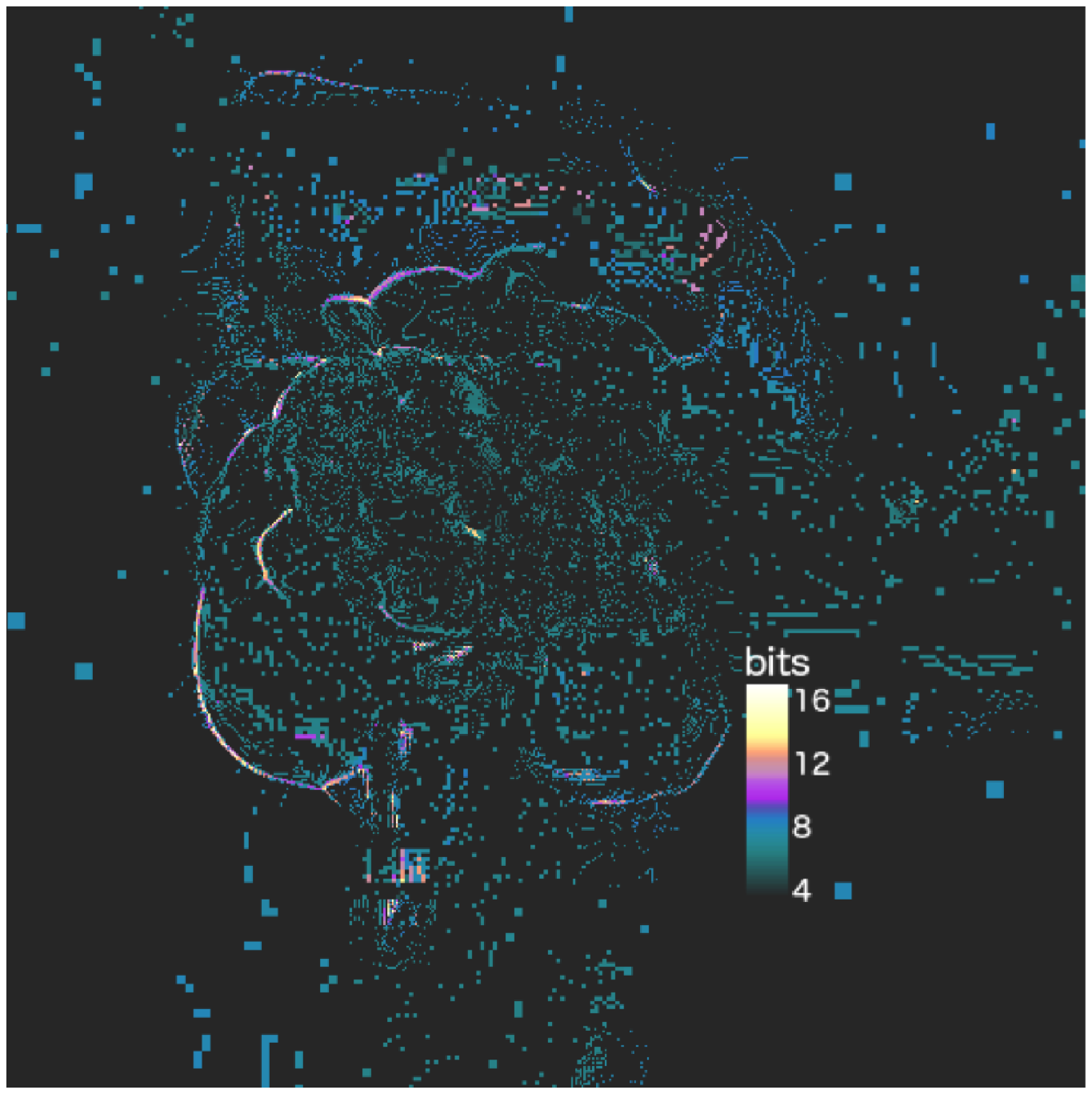}
\includegraphics[width=0.33\textwidth]{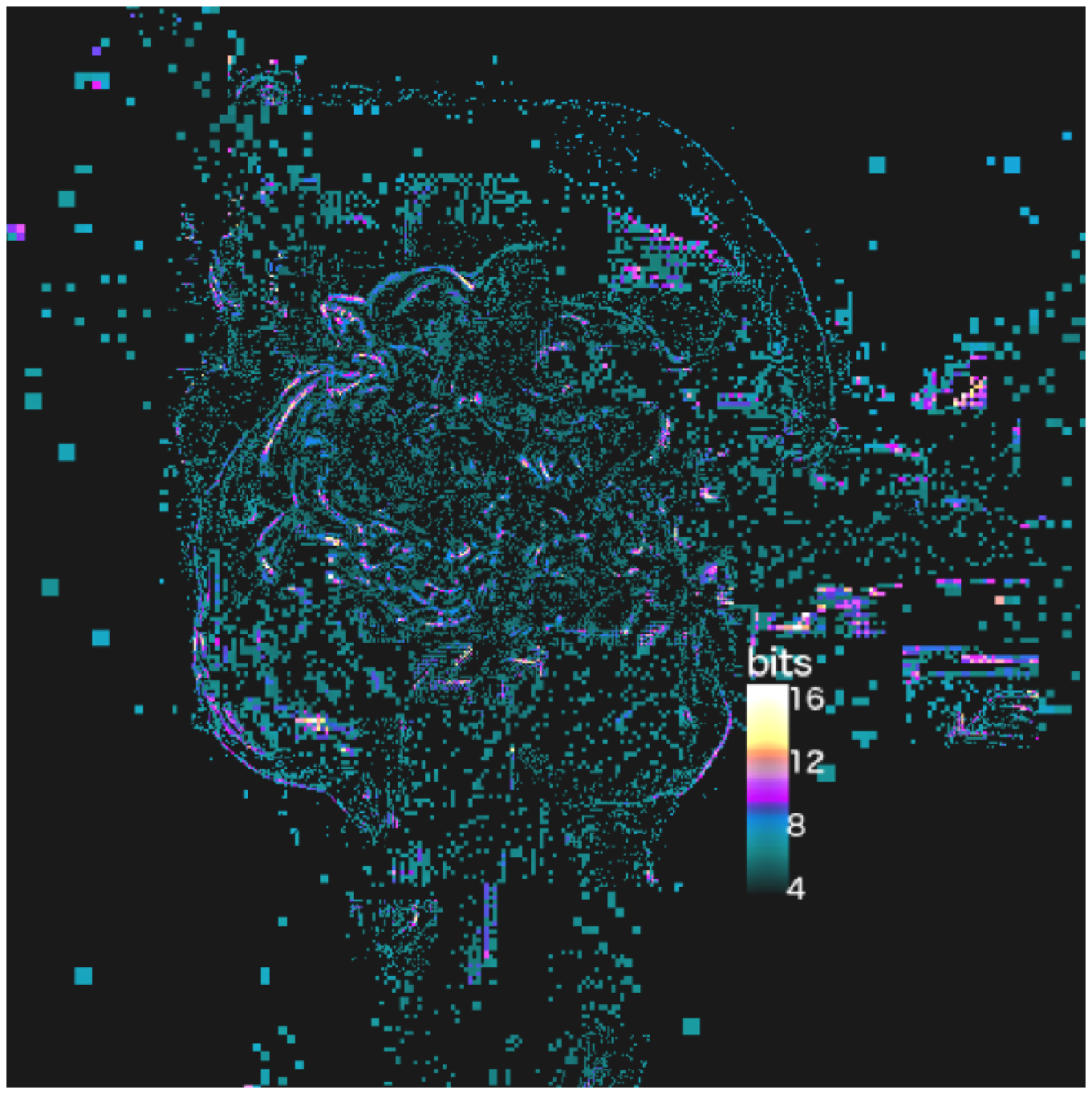}
\caption{For the same selection of Fig.\ref{fig:map0}, shown are the maps of statistical complexity (in unit of bits) for the kinetic, thermal and magnetic energy components, respectively.}
 \label{fig:map1}
\end{figure*}

%\section{Results}
%\label{cosmo}

\subsection{Cosmological simulations of galaxy clusters}
\label{cosmo}

The most important analysis of this work concerns the study of the formation of galaxy clusters in cosmology.
A large variety of algorithms are available for numerical cosmology, including either particle-based or grid-based methods to couple the evolution of the gas and dark matter component \citep[e.g.][]{do09,bk11}, often yielding a promising convergence in code cross-comparisons  \citep[e.g.][]{1999ApJ...525..554F,2008CS&D....1a5003H,va11comparison,2012MNRAS.423.1726S}.  

In this work, we based on the Eulerian representation of gas physics in the expanding space-time, given by the cosmological code {\enzo} \citep[][]{enzo14}, which we already used in many works \citep[][]{va11turbo,va11entropy,va14mhd}. {\enzo} is a highly parallel code for cosmological (MHD)hydro-dynamics, which uses a particle-mesh N-body method (PM) to follow the dynamics of the DM and a variety of hydro-MHD solver to evolve the gas component on a support uniform or adaptive grid \citep[][]{enzo14}.  
Our simulations include the effect of magnetic fields, radiative cooling of gas and energy feedback from active galactic nuclei. 
Adaptive mesh refinement (AMR) was used to selectively
increase the dynamical resolution in the formation region of galaxy clusters, which is mandatory to properly resolve magnetic field amplification \citep[][]{xu09}. We used the same set of initial conditions and cosmological parameters 
in \citet{va11turbo} to resimulate one galaxy cluster with a total mass of $1.12 \cdot 10^{15} M_{\odot}$ and a virial radius of $R_{\rm vir}=3.2 ~\rm Mpc$ at $z=0$.  We started from a volume of $260^3$ Mpc$^3$ (comoving) with an initial root grid of $256^3$, additionally refined 5 times (with a $\times 2$ refinement, up to a maximum resolution of $31.7$ kpc) inside a sub volume of $\sim 25^3$ Mpc$^3$ centered on the cluster. In this work, we use an aggressive AMR strategy and refine the grid wherever local overdensities $\geq 10\%$ than of surrounding are found, as well as whenever velocity jumps larger than $\leq 1.5$ are detected. This ensures that typically $\sim 80\%$ of the cluster volume is refined up to highest resolution, which ensures a large enough dynamical range to follow the turbulent accretion flows within the cluster \citep[][]{va11turbo,va14mhd}. More details on this simulation can be found in \citet{wi16}.

To better disentangle gravitational and non-gravitational evolutionary effects on the cluster evolution, we resimulated this object twice: a) only including gravity, hydrodynamics and magnetic fields and b) additionally including radiative gas cooling and thermal/magnetic feedback from active galactic nuclei. In the second case we allowed the release of thermal energy ($10^{60}  \rm erg$ per event, starting from $z=4$) and magnetic energy ($10^{59} \rm  erg$ per event, as a dipole structure) at the location of high density peak within the cluster volume.  This simplistic modeling of feedback active galactic nuclei bypasses the problem of following prohibitively small scales involved in the accretion of gas onto supermassive black holes, but allows us a correct descriptions of the interplay of cooling and feedback in these simulations {\footnote{This simplified approach can properly reproduce the thermodynamical properties of the observed ICM on $\geq 100 ~\rm kpc$, as shown in our previous works \citep[][]{va11entropy,va13feedback,va16scienzo}}}. We initialized the magnetic field uniformly in the volume of both runs, to the comoving constant value of $10^{-10} \rm G$.
For the following analysis of complexity, we saved $\sim 440$ snapshots of both simulations, by writing all physical fields with a constant time spacing of $\Delta t \sim 3.11 \cdot 10^{6}$ yr.  We give the complexity analysis of the non-radiative run in Sec.\ref{subsec:non-rad}, while we study the extra-complexity resulting from non-gravitational effects in Sec.~\ref{subsec:rad} \\

Fig.~\ref{fig:map0} shows the spatial distribution of thermal, kinetic and magnetic energy for a slice crossing the center of the cluster in our non-radiative simulation at $z=0$. 
While the kinetic energy dominates the gas infall regions outside of the cluster volume, the thermal energy is dominant within the cluster as a result of the thermalization of infall kinetic energy via shock dissipation, starting at the outer strong accretion shocks \citep[e.g.][]{ry03}. The kinetic energy budget is however still significant within the cluster, owing to residual subsonic turbulent motions \citep[e.g.][]{va11turbo,mi15}. 
The magnetic energy is small everywhere, and only in localized patches it reaches a few percent of the thermal/kinetic energy within the cluster, while it is $\sim 10^{-4}$ in most of the volume. However, the magnetic energy can be significant compared to the  thermal energy in supersonic flows where the gas thermalization is inefficient, i.e in cluster outskirts and within filaments connected to the cluster. {\footnote {It shall be noted that simulating the growth of ICM magnetic fields in a small-scale dynamo is still a challenge, owing to the limited dynamical range which can be reached even with high-resolution numerical simulations \citep[e.g.][]{va14mhd,2016ApJ...817..127B}. For this reason, we the magnetic field level reached in our runs is lower than suggested by observations \citep[e.g.][]{bo13}.}

\begin{figure}
\includegraphics[width=0.495\textwidth,height=0.9\textheight]{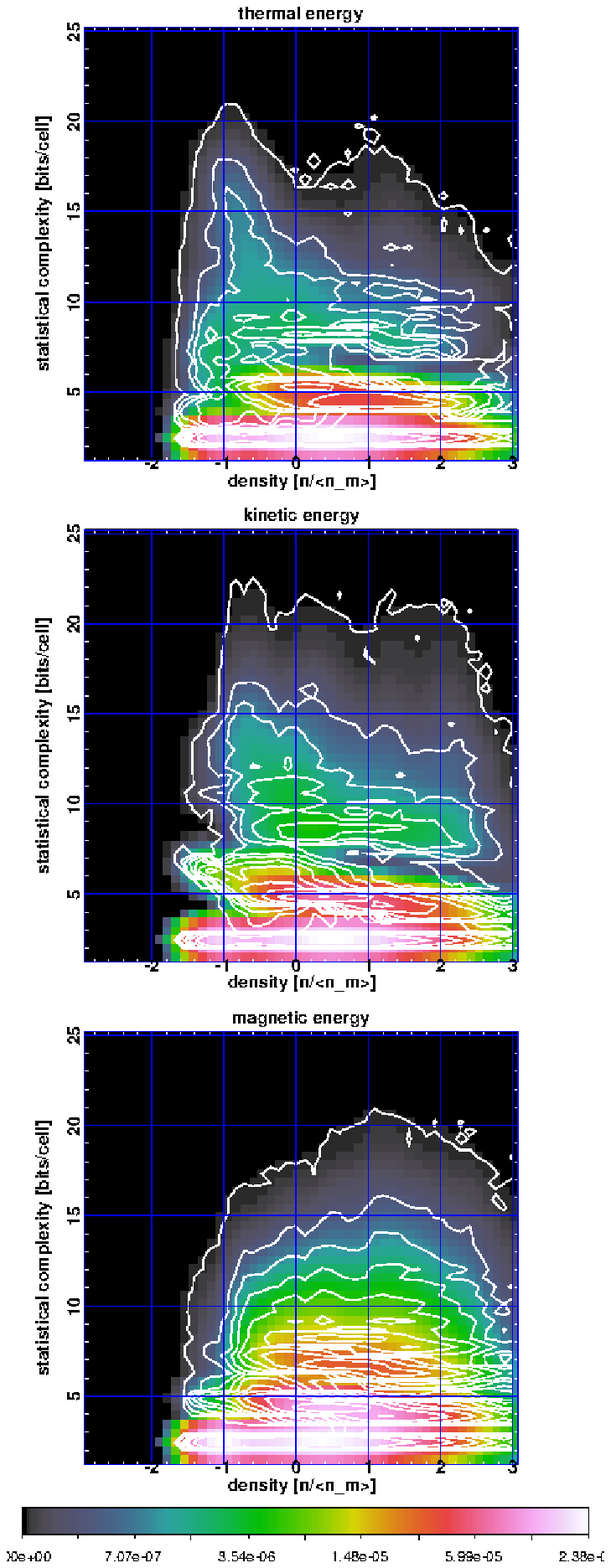}
\caption{Phase diagram comparing the statistical complexity of the thermal (top), kinetic (center) and magnetic (bottom) energy as a function of the gas density (relative to the total mean density. The color coding gives the volume fraction of cells in a certain phase.}
 \label{fig:phase}
\end{figure}

\subsubsection{Statistical complexity in the non-radiative run}
\label{subsec:non-rad}

The methods outlined in the previous Sections allow us to compute the cell-wise complexity in the simulation and to treat this as an additional derived 3-dimensional field characterizing the simulated ICM. 
The statistical complexity for each cell, $C_{\mu,xyz}$ (in units of bits per cell), for the simulated cluster is given in Fig.~\ref{fig:map1} for each energy field separately, and in the false color image in Fig.~\ref{fig:map_rgb} for the whole the fields combined. 
To achieve a very refined description of the cluster volume, we employed here $N_{\rm bin}=200$ logarithmic energy bins and compared two snapshots separated by  $\Delta t  \approx 7.3 \times 10^7$ years (i.e. a timestep at the root grid level). 

The  maps of statistical complexity show at the same time more finer details than the input energy fields, and also noisy structures of clear numerical origin. The time-based analysis of statistical complexity is designed to focus on fast energy fluctuations at the cell scale, and to highlight those that are more difficult to predict based on the past evolution. In this respect, the statistical complexity filter is very efficient to highlight, both, the signature or physical processes such as  shocks and turbulent motions, as well as the small-scale numerical noise associated to our aggressive mesh refinement strategy  \citep[e.g. see discussion in ][]{sch15}.

However, most of the ``complexity patterns'' inside the cluster are manifestly associated with physical jumps of the energy fields, with each energy fields displaying its distinct complexity pattern.
The complexity of the kinetic and thermal energy is mostly increased in narrow zones connected to shocks. 
Both $E_T$ and $E_K$ are modified by shock jump conditions on this short time scale. $E_T$ requires $\sim 5-10$ times more information because at strong non-radiative shocks the jump of thermal energy is much larger than that of kinetic energy. 
Although each shock is described by the ``simple'' Rankine-Hugoniot jump conditions, at every timestep only a small fraction of the cells in a given energy bin is crossed by shocks. Hence additional information is needed to predict the occurrence of shocks in cluster outskirts at any given time, of the order of $\geq 10 ~\rm bits/cell$ at this redshift. It is worth recalling that in the framework outlined in Sec.~\ref{methods}, the notion of complexity describes the amount of information necessary to predict the evolution of the system at a given time/spatial location. In this respect, predicting the occurrence of shocks in the ICM depends on a number of factors, which requires a significant
amount of information to compute, even if the underlying physics is ``simple''.
 Compared to the thermal energy, the kinetic energy $E_K$ is complex in a larger volume fraction and also closer to the cluster center. The ICM is known to host volume filling subsonic turbulence at all epochs, as result of gravity-driven random motions \citep[e.g.][]{va11turbo,mi15,va16turbo}. The kinetic energy is thus subject to complex fluctuations due to turbulence even on the $\approx 7.3 \times 10^7 \rm yr$ timescale. 
The magnetic energy, $E_B$, outnumbers the complexity budget of the other two fields, by $\sim 10-100$ times.
The magnetic energy is sub-dominant compared to the thermal/kinetic energy of the ICM, which means that the magnetic field lines are continuously subject to the violent stirring gas motions, which drive fluctuations of the magnetic energy on short timescales. Moreover, the predominantly
solenoidal turbulent motions are responsible for small-scale dynamo amplification of ICM magnetic fields over time \citep[e.g.][]{xu09,va14mhd,2016ApJ...817..127B}, which makes the evolution of the magnetic energy even more complex. \\

A more systematic view of the distribution of {\bf $C_{\mu,xyz}$} as a function of the gas density (normalized to the mean matter density, $n_M$) is given in the phase diagram on Fig.\ref{fig:phase}. While the complexity associated to the thermal gas energy is larger at the low gas over-density typical of outer accretion shocks ($n/n_M \sim 0.1-1$), the complexity of the kinetic energy has a distribution which extends more towards higher over-densities. Finally, the magnetic energy has a peak of complexity extending to $n/n_M \sim 10-10^2$, i.e. in the innermost cluster regions where the stirring by turbulent motions is more volume filling.

It is worth stressing once more the powerful capabilities of complexity analysis. 
The symbolic analysis of the datastream can identify the most complex pattern arising from the hydro and MHD dynamics of the 
system, which are in great majority shocks and turbulent motions in this case. Standard approaches to identify these important ingredients
for the evolution of the ICM require resorting to ad-hoc numerical filters or finding schemes, specifically tailored to identify 
shocks  \citep[e.g.][]{ry03,va11comparison} or disentangle laminar from turbulent components \citep[e.g.]{va12filter,mi15,va16turbo}.
Instead, all complexity patterns found by our algorithm are the result of an entirely symbolic analysis of the datastream, i.e. the patterns are recovered with high accuracy just by comparing the statistics of energy transitions $E_j(t+\Delta t)|E_i(t)$  between two timesteps. In a sense, this operation is performed {\it blindly} over the data, meaning at no level there is a physical description of the underlying dynamics or physical laws that the system
is subject to. This means that in general the complexity filter has the potential to unveil interesting flow patterns even in unexpected regions, which can be neglected in the absence of proper numerical filters. %The following Section shows indeed that Information Theory can highlight the emergence of complexity from non-gravitational effects, which can be lost by more standard analysis.

\begin{figure*}
\includegraphics[width=0.99\textwidth]{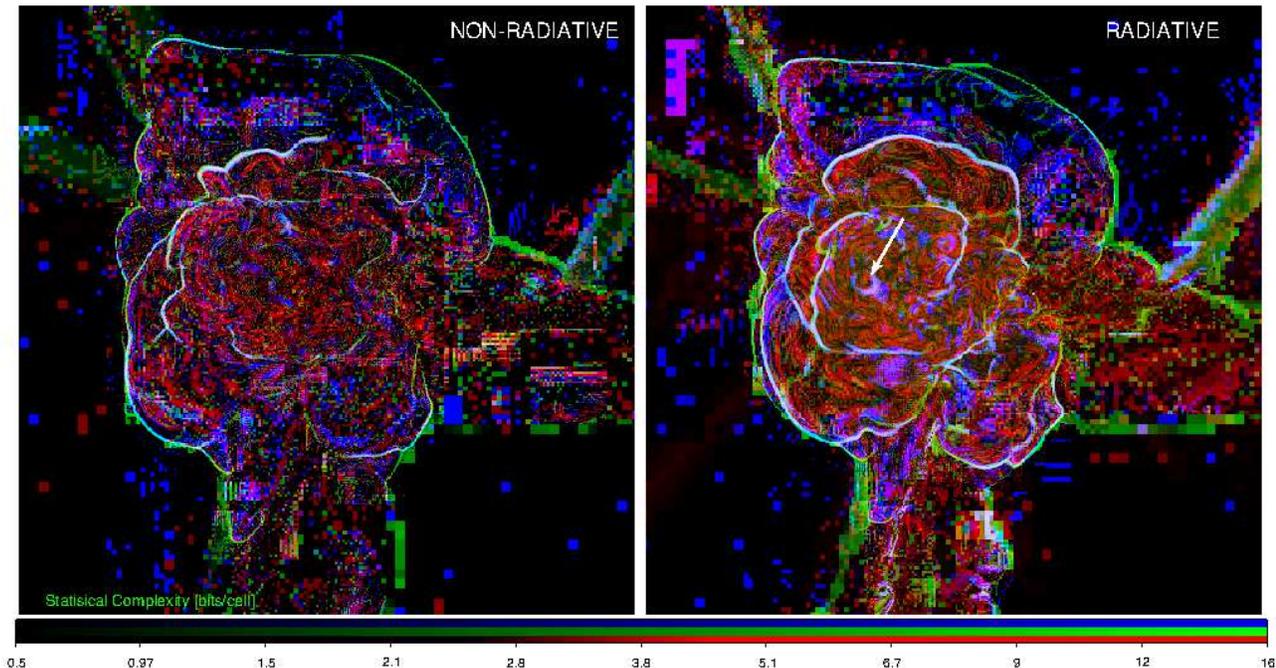}
\caption{False color rendering of the complexity (in $\rm bits/cells$) of our simulated cluster in the non-radiative (left) and in the radiative (right) setup. The red color show the magnetic complexity, the green the thermal complexity and the blue the kinetic complexity. The white arrow in the right image mark the location of a central AGN in the radiative run.}
 \label{fig:map_rgb}
\end{figure*}

\begin{figure*}
\includegraphics[width=0.949\textwidth]{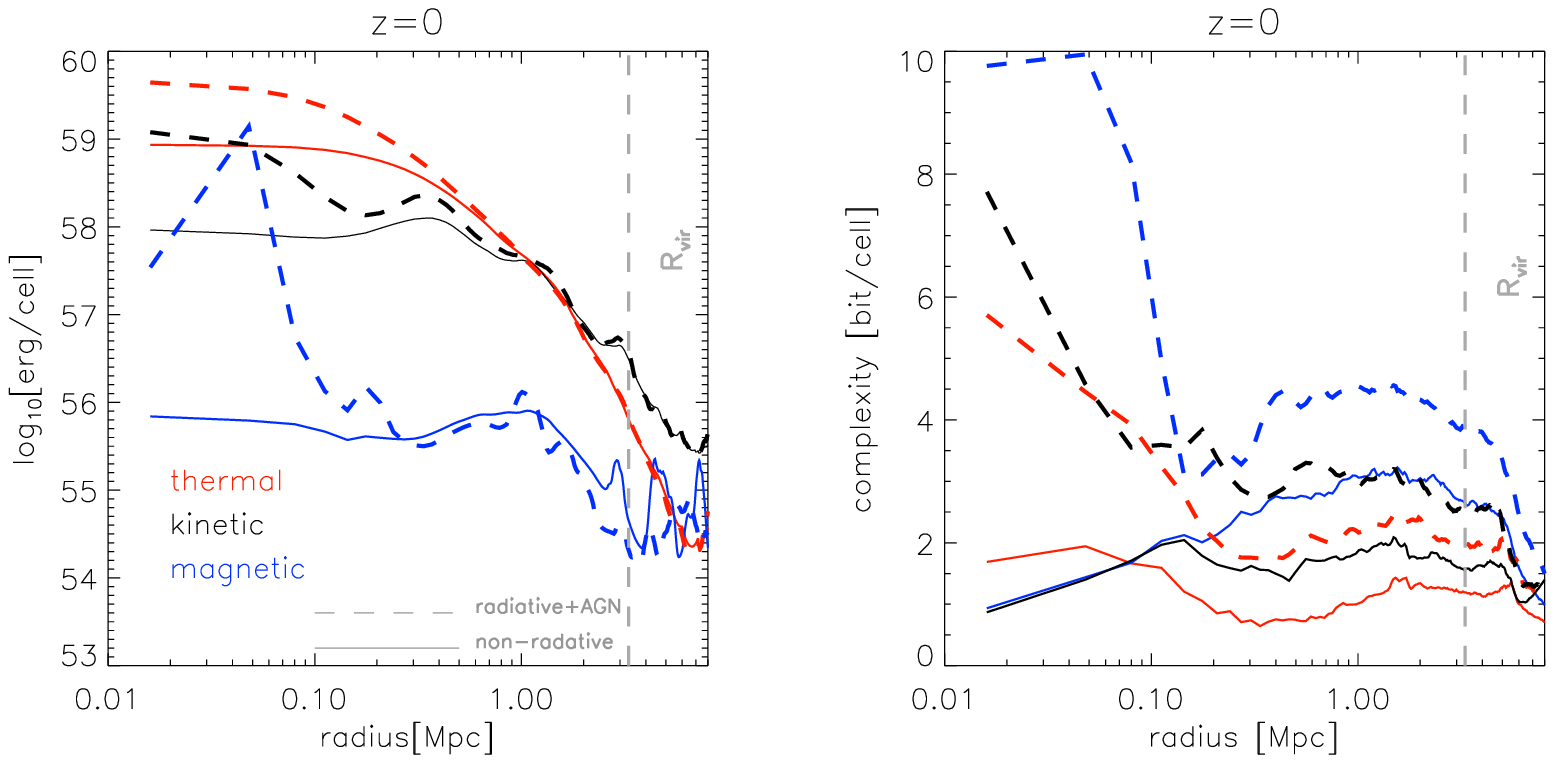}
\includegraphics[width=0.949\textwidth]{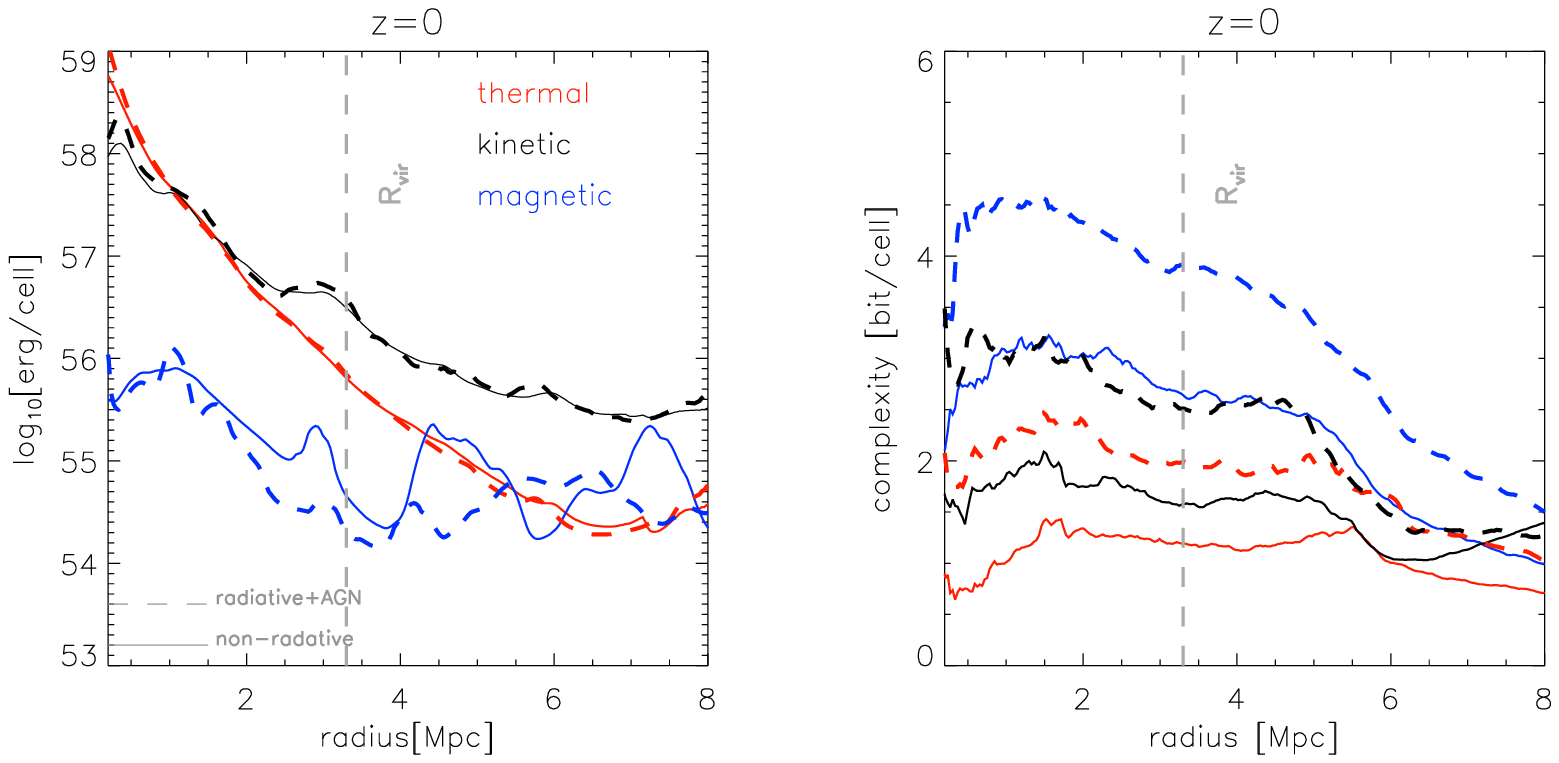}
\caption{Radial profiles of the thermal, magnetic and kinetic energy (left) and of the statistical complexity (right) for our simulated cluster at $z=0$, for the non-radiative (solid lines) and the radiative (dashed lines) run. The top panels use a logarithmic stretching of the x-axis to better highlight the profile in the cluster core.}
 \label{fig:profiles}
\end{figure*}

\subsubsection{Complexity from gravitational vs non-gravitational effects}
\label{subsec:rad}

The application of Information Theory can highlight the additional role of non-gravitational physics in generating complex evolutionary patterns in the simulated ICM. 

The false color maps of Fig.\ref{fig:map_rgb} compares the distribution of statistical complexity in the non-radiative (left) and in the radiative (right) run, for a one-cell thick slice through the cluster center at $z=0$. This comparison well highlights, both, the large-scale similarities of the two runs and the additional differences due to non-radiative physics.
In both cases, the magnetic complexity (in red) is always found to dominate within the cluster volume, with maxima 
downstream of merger shock waves moving outwards. The kinetic complexity is second in relevance. It is more closely associated to shock jumps than the magnetic complexity, and appears to be dominant in the outer shell approaching accretion shocks.
Finally, the thermal energy has prominent maxima of complexity at accretion shocks, marking the sharp transition between the smooth and the structured gas at the periphery of the cluster. Similar patterns are observed also within filaments connected to the simulated cluster (i.e. right and bottom sector of the image). 
The radiative simulation displays a $\sim 5-10\%$ higher level of complexity in all energy fields, across most of the cluster volume.
The relative trend between the 3 fields is similar, yet in the radiative run the magnetic complexity is found to be more volume filling and to
extend more towards cluster outskirts (e.g. lower half of the cluster volume in Fig.\ref{fig:map_rgb}).
The extra-complexity in the radiative run is also evident in the proximity of an active AGN source in this slice, which we mark with a white arrow  in the right panel. Close to these regions, the powerful output of thermal/magnetic energy and the outflow it drives produces patterns of 
complexity that are not present in the non-radiative simulation.\\

The relative trend of complexity in the two runs is visualized also by the radial profiles of Figure \ref{fig:profiles} (with or without a logarithmic stretch of the x-axis, to alternatively focus on the trends in the core or in the outskirts of the cluster), together with the radial profiles of the energy fields.
As observed above, cooling and AGN feedback have a strong effect on the energetics of the cluster core, and on its complexity level as well.
In the innermost $\sim 300 \rm ~kpc$ of the radiative simulation, both the thermal and the kinetic energy are $\sim 2-5$ times larger as a result of the increased gas density and of the extra heating from the AGN. The magnetic energy is even $\sim 10^2$ larger than in the non-radiative case, because the extra magnetisation from AGN dominates the contribution from primordial fields. 
However, on scales larger than $\sim \rm Mpc$ the energy profiles of the radiative run at are extremely similar to the non-radiative case. 
Cooling is increasingly less significant at lower gas densities, and simulated sources of feedback have overall a small impact on very large volumes. 
This is radically different in the case of the complexity profiles, even outside of the virial radius of the cluster, consistently with the 
what we saw already in Fig.\ref{fig:map_rgb}. 

%At the physical level, this is expected because each energy fields evolves according to additional 
%physical mechanisms in addition to gravity and hydrodynamics - this is also easily understood at the numerical level, for the presence of 
%additional source (i.e. AGN feedback) and loss (i.e. cooling) terms in the hydrodynamical equations solved by the code at every timesteps. 

Close to the central source of feedback in the cluster core (white arrow in Fig.\ref{fig:map_rgb}) the additional thermal and magnetic
energy from the AGN dominates the energy evolution on the root grid timescale, leading to a $\sim 10$ times larger complexity for the
magnetic energy, and to a $\sim 3-5$ larger complexity for the other two energy fields. Even at larger distances from the central AGN, the complexity is still on average larger by $\sim 50-100\%$ compared to the non-radiative run. Although the energy profiles (left panels) are quite similar at large radius, in the non-radiative case the profiles is the result of a global balance of cooling losses, enhanced compression and AGN feedback (from the central cluster AGN and but also from several others associated to substructures accreted by the main cluster), which makes the global evolution of the ICM more complex at all radii.
At the physical level, this is expected because each energy fields evolves according to additional 
physical mechanisms in addition to gravity and hydrodynamics - this is also easily understood at the numerical level, for the presence of 
additional source (i.e. AGN feedback) and loss (i.e. cooling) terms in the hydrodynamical equations solved by the code at every timesteps. \\

Where did these differences in complexity originate? 
The block entropy analysis introduced in Sec.~\ref{subsubsec:block_entropy} allows us to monitor how complexity has grown over the cluster lifetime. In addition, we can relate the emergence of complexity to specific events that affected the global energetics of  the ICM at different redshifts. 
To compute the block entropy, $H(L)$, and its source rate term, $h_\mu(L)$, we analyzed the evolution of the $E_K$, $E_T$ and $E_B$ energy fields from $z=30$ to $z=0$ within a fixed reference volume. For each simulated cell we computed the symbol statistics as a function
of the increasing length $L$ of the data stream, using $N_{\rm bin}=5$ logarithmic energy bins. 
We remark that a complete analysis of the full simulated sequence of symbols in the cluster volume, $X(L)$,  is made challenging by the enormous amount of data which is required: following $\sim 800^3$ high resolution cells over $\times 440$ time steps, even by binning the energy values in $5$ energy bins, approximately require to keep in memory $\sim 1.8$ Tb of data, for each energy field separately. 
This is prohibitive and in this first exploratory work we restrict ourselves to computing the block entropy of a smaller dataset which still can give a representative sampling of the evolution of cosmic gas. In detail, 
we used a 2-dimensional selection of  $440 \times 440 \times 1=193,600$ cells through the cluster center, corresponding to a $15.2 \times 15.2 ~\rm Mpc^2$ region, and sampling the datastream every 5 root grid timesteps ($\sim 3.6 \times 10^8 \rm yr$). 
The  choice of such selection is motivated by the fact that, given a maximum number of cells to follow, a wide 2-dimensional selection can better statistically follow, both, the evolution of the central cluster region as well as of the cluster outskirts, which are equally important for the growth of complexity. \\

The top panel of Figure \ref{fig:block_entropy} gives the evolution of the integrated values of  $E_K$, $E_T$ and $E_B$ , as well as of the enclosed gas mass (which is about $\sim 1/6$ of the total gas+dark matter mass), for such 2-dimensional selection.   
The evolution of the non-radiative run (solid line) is here contrasted to the evolution of the radiative run with feedback (dot-dashed line). 
Although the global evolution of the energy fields is quite similar, the usual differences appears:  a) in the radiative case, $E_T$ is always lower than in the non-radiative case, due to the loss of energy via cooling. This effect is particularly important in the first evolutionary stage $\leq 4 \rm ~Gyr$, while after this the energetics is dominated by the heating of infall kinetic energy and then it slowly approaches the same
level in both runs. In addition, AGN are compensating the radiative losses via thermal feedback, at least globally in the ICM volume. b) The magnetic feedback by AGN and the extra compression by cooling increase the magnetic energy within the volume compared to the non-radiative case. By the end of the simulation the final magnetisation level get again comparable in both cases, meaning that the magnetic field in the volume is dominated by the amplification of primordial fields. \\

The differences in block entropy are more spectacular.
As expected, $H(L)$ increases in a monotonic way and flattens over time, reaching a maximum of $H \sim 12-16 \rm ~bits/cell$ by the end.  In general, we observe that the block entropy of all energy fields is sharply increased in correspondence of all important mergers and matter accretions experienced by the cluster ($t \sim 4$, $\sim 6$, $\sim 9$, $\sim 10$ and $\sim 12$ $\rm Gyr$, see gray lines in the Top panel). However, at early times the cooling-feedback loop adds significant block entropy to the gas in the radiative run.
In particular, for $t \leq 6-7 ~\rm Gyr$ ($z \geq 1$) the block entropy of all fields in the radiative runs is already significantly larger than in the non-radiative setup.  At this epoch the medium is far from virialization, and the balance of 
 cooling and AGN feedback dominates the energetics of ICM on small spatial and time scales.

Although the final block entropy level of $E_T$ is similar in the two setups ($\approx 15$ versus $\approx 16 ~\rm bits/cell$ when the non-radiative and radiative runs are compared), these complexity levels have been produced at very different epochs.  
For example, in the non-radiative case, $E_T$  has reached $90\%$ of the final block entropy level at $t \approx 11 ~\rm Gyr$, while
in the radiative case this level has been already reached by $t \approx 7~\rm Gyr$.
The reason of these differences is better highlighted by the entropy gain, $h(L)$, which computes the increase of the block
entropy as function of the symbol length $L$ (in this case, the elapsed time). 
The lower panel of Fig.\ref{fig:block_entropy} shows that 
the "extra" complexity in the radiative run is acquired very early, $t \sim 1-5 ~\rm Gyr$, i.e. before the cluster assembled. The maximum of $h(L)$ is found at $t \sim 2 ~\rm Gyr$, i.e. close to $z=4$ epoch which marks the begin of the AGN feedback in our numerical setup (Sec.\ref{cosmo}). The other energy fields ($E_B$ and $E_K$) display a maximum production
of block entropy with a delay of $\sim ~\rm 2 Gyr$. We link this to the outflows driven by AGN, which is observed to drive turbulence and magnetic field amplification in excess to the non-radiative case at the same epoch \citep[][]{va13feedback}.

In a second stage, we observe a second  
significant peak of entropy gain in all energy fields ($\sim 1.5-2.5 ~\rm bits/cell$), after the bulk of the cluster mass has been assembled and the virialisation process is still ongoing. This shortly follows to a sharp mass increase experienced by the cluster at this epoch ($t \sim 6$ $\rm Gyr$), which is marked by a peak in the total mass distribution (upper panel). 
At this epoch, all energy fields in the non-radiative run are significantly less complex than in the radiative case, and the chaotic motions following merger events add proportionally more complexity to the ICM. 

While the absolute value of block entropy at a specific epoch is dependent on the specific choices of the binning of energy levels and on the time sampling frequencies (see Appendix), the relative growth of  block entropy in the energy fields is robust to model variations.
We give in Fig.\ref{fig:block_ratio} the ratio between the block entropy of each field and the total block entropy of each run, which highlights the shift in the relative complexity of the various field, when radiative physics is included. Cooling and feedback have overall a little impact on the relative complexity of the energy fields after the cluster assembled, $t \geq 4 ~\rm Gyr$. The role of cooling and feedback
is more marked at earlier times.  For example, in the radiative case we observe a larger relative
importance of the magnetic and kinetic complexity as an affect of gas compression and outflows released by the onset of AGN feedback. This stresses once more that, longer before contributing to the mass of the $\sim 10^{15} M_{\odot}$ cluster that will dominates this region at late redshift, the cosmic gas in a realistic simulation has been subject to a very complex evolution in its thermal, kinetic and magnetic properties.
In the Appendix, we also present additional tests showing how the above trends are robust against sampling variance, i.e. if different slices through the cluster volume are used to measure the growth of block entropy. While some scatter is present, the variations
are in general much smaller than the relative difference in block entropy of the different fields.

\begin{figure}
\includegraphics[width=0.49\textwidth]{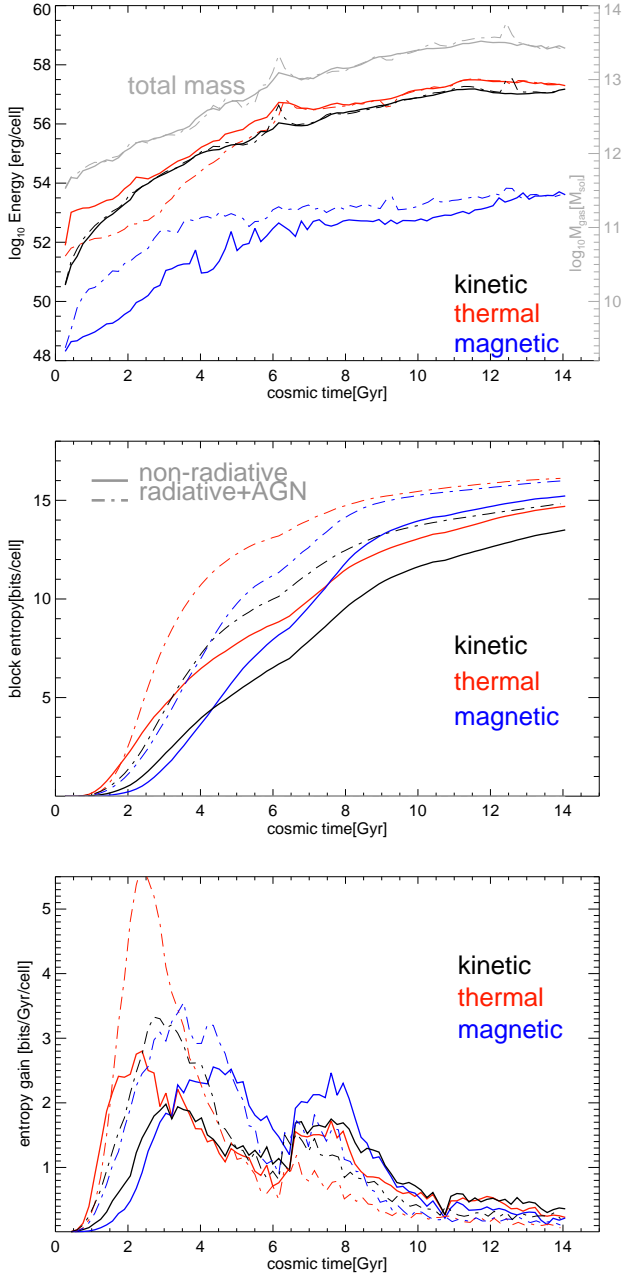}
\caption{Top panel: evolution of energy fields and of the total gas mass (gray lines), for a selection of $193,600$ cells in the central region of our runs. Central panel: evolution of the block entropy and of the total gas entropy for the same selection of cells. Bottom panel: evolution of the entropy gain for the same selection of cells. In all panels, the solid lines show the trends of the non-radiative run, while the dot-dashed lines are for the radiative run with feedback.}
 \label{fig:block_entropy}
\end{figure}

\begin{figure}
\includegraphics[width=0.49\textwidth]{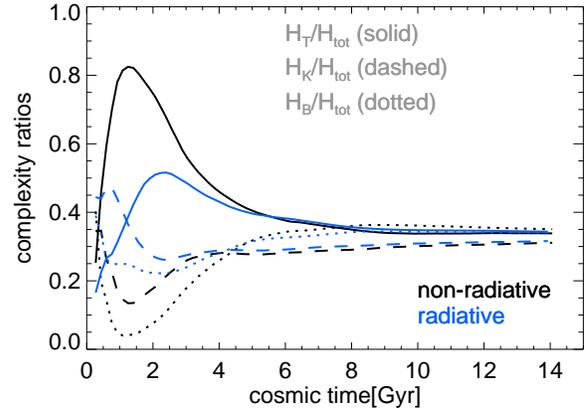}
\caption{Evolution of ratio between the block entropy of the different energy fields and the total block entropy, for the same region of 
Fig.\ref{fig:block_entropy}. The black lines are for the non-radiative run while the blue lines are for the radiative run.}
 \label{fig:block_ratio}
\end{figure}

\begin{figure}
\includegraphics[width=0.49\textwidth]{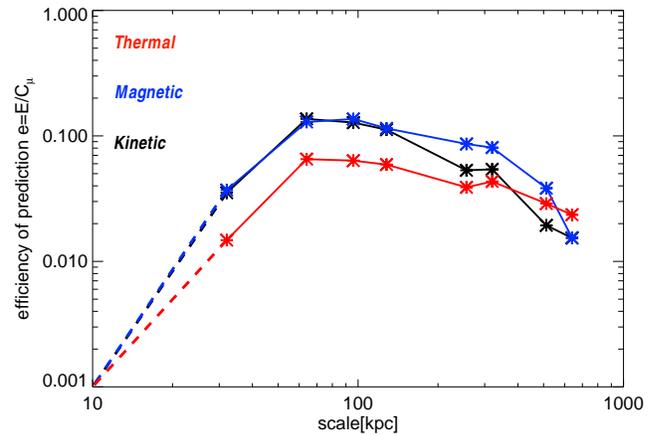}
\caption{Relation between the efficiency of prediction (Sec.\ref{subsubsec:eff}) at different interpolation scales, for the same
2-dimensional selection of Sec.\ref{subsec:rad}. The dashed lines connects the values of $e$ measured in the simulation, with the $e$ values estimated from plasma physics on unresolved scales (see text for explanation).}
 \label{fig:efficiency}
\end{figure}

\subsubsection{Efficiency of prediction: which scales are best to predict the evolution of the ICM?}
\label{eff_icm}

The volume comprised by galaxy clusters is so large that it is possible to study plasma processes of the ICM 
on many different scales. 
Following Sec.\ref{subsubsec:eff} we investigate which is the best scale to produce a predictive model for the ICM, based 
on the efficiency of prediction, $e=E/C_\mu$, which we measured by interpolating our simulated fields
on coarser scales. In detail, we computed $H(L)$, $E(L)$ and $C_\mu$ for the same 2-dimensional slice of  $440 \times 440 \times 1$
cells through the cluster center used in  Sec.\ref{subsec:rad}, after linearly interpolating the energy fields on increasingly
coarser resolutions, from the original resolution of $31.7$ kpc up to $634$ kpc (20 cells). 
Through this procedure it is possible to investigate at which scale the self-organisation of the ICM is most emergent, and its
evolution most efficient to compute. For simplicity,  we  limit our analysis to the non-radiative run, and consider the efficiency of prediction at $z=0$, using $N_{\rm bin}=10$ logarithmical energy bins to compute $E(L)$ and $C_\mu$
Figure \ref{fig:efficiency} shows the trend of $e$ measured in our simulation as a function of scale, which displays a similar
behaviour for all energy fields. The maximum efficiency of prediction is somewhere in the range $\sim 63-95$ kpc, with $e \approx 0.15$ for the kinetic and magnetic energies, while $e \approx 0.08$ for the thermal energy{\footnote{It shall be noticed that even the maximum measured $e$ is quite far from the theoretical maximum of $1$. However, considering the number of approximations in our modelling and the somewhat limited volume that we can presently analyse with these algorithms, the estimates given here are expected to give at least a robust relative trend of $e$ with scale.}}. While predicting the evolution
of the kinetic and magnetic energy is more efficient for $\leq 300$ kpc, on larger scales the thermal energy offers the
largest efficiency of prediction. 
These result can be interpreted by noticing that the scale at which the efficiency of prediction of the simulated ICM is maximum is tentatively close to the typical scale of turbulent eddies in the simulated ICM \citep[e.g.][]{va11turbo,va12filter, mi15,va16turbo}, to the typical correlation scale of observed and simulated magnetic fields \citep[e.g.][]{xu09,bo13}, as well as to the scale of measured projected density fluctuations in X-ray \citep[e.g.][]{zhur15}. Therefore, it is reasonable that $e$
gets maximum on scales where the ICM flows have the largest degree of dynamical organisation. 

As an important caveat to our analysis, it shall be noted that in general the entropy gain (Eq.\ref{gain}) is systematically underestimated when the sequence of symbols (and hence the block length) is large. Together with a more extensive tests of the efficiency of prediction with a larger dataset of objects (including different dynamical histories), with future tests we will assess the dependece of this preliminary result on the finite size of the analysed sample of data.

The limited numerical resolution of our runs prevent us to directly compute the efficiency of prediction for
smaller scales. However, the classic hydro-MHD picture of the ICM must break for $\ll ~\rm kpc$ scales, in a regime where
wave-wave, particle-wave and particle-particle interactions are important \citep[e.g.][]{sch05,bl11}. 
In this regime, to estimate the efficiency of prediction
 at such  "microscopic" scales we can resort to the same arguments by \citet[][]{2004PhRvL..93n9902S} and \citet{prokopenko2009information}, adapted to the ICM conditions. The dynamics of particles in the ICM can be assumed to be Markovian at first order (i.e. the thermodynamical value of a single particle only depends on the last microstate), hence $E=C_\mu-L h_\mu \approx C_\mu-h_\mu$ because $L \approx 1$. 
In a perfect gas, the thermodynamic entropy gives the statistical complexity, and for the thermal particles in the ICM this is $S \sim 10 \rm keV/particle$ for a $M \sim 10^{15} M_{\odot}$ cluster \citep[e.g.][]{bk11}. The entropy rate crucially depends on the mode of energy/entropy exchange between particles on short timescales, which drastically diverge if different models of the ICM are assumed.
Two extreme scenarios can describe the exchange of energy (and hence information) between particles of the ICM \citep[e.g.][]{SA88.1,sch05,2011MNRAS.410.2446K,bl11}: a classic {\it collisional} view in which Coulomb collisions between thermal particles are the channel to exchange energy (with a typical collision time of $\sim 3.3 \cdot 10^{5} \rm yr$ for electron-electron collisions and $\sim 1.4 \cdot 10^{7} \rm yr$ for proton-proton collisions), or a more realistic{\it weakly collisional}  view, in which energy is exchanged via the mediation of collective plasma effects (with enormously smaller timescales, $\sim $ seconds). 

In the first scenario, from the proton-proton Coulomb collision frequency we can estimate an entropy rate of $h \approx 10^{-7} \rm keV/particle/yr$, which implies that the efficiency of prediction approaches unity only at very small timescales (i.e. $e \sim 0.99$ for $\leq 10^{5}$ \rm yr) while for astronomically relevant timescales it drops to zero. 
In the second scenario, the extremely fast action of plasma collective implies that on microscopic scales the efficiency of prediction is $\approx 1$ only in the scale of {\it seconds}, while it rapidly drops to zero for any other longer timescale.\\

This exercise quantitatively shows the obvious fact that a detailed thermodynamical view of single particle interactions in the ICM is irrelevant to predict the evolution of the ICM on astronomically relevant scales, given the enormous difference in scale between microscopic and macroscopic processes involved {\footnote {However, we notice that in the case of fast growing instabilities the above picture might further change \citep[][]{2011MNRAS.410.2446K}, and consequently the efficiency of prediction might have a different trend with scale.}}.

\section{Discussion and Conclusions}
\label{conclusion}

In this work we presented the first application ever of Information Theory to the study of cosmic structures.
We succeed in quantifying the complexity associated to the formation of galaxy clusters and to relate its growth to
the dynamical evolution of the ICM gas as a function of time and in response to different physical mechanisms.

In summary, our study shows that:

\begin{itemize}

\item the algorithms from Information Theory implemented in Sec.\ref{info} can easily detect the emergence of hydrodynamical structures
in the simulated ICM, just based on the symbolic analysis of the output of the simulation and {\it without knowing anything about the underlying dynamics}. This is possible because complexity analysis can read out the emergence of complex behaviors directly from the symbolic data stream generated by the simulation.

\item Shocks and turbulent motions are very well captured by this analysis, as well as the footprints of AGN activity in the ICM. Beside identifying ``expected'' important hydrodynamical features in the ICM, complexity analysis has the potential to unveil {\it unexpected} complex pattern in simulations (which might also be of spurious numerical origin).
%, because it is just based on the analysis of the symbolic datastream produced by the simulation code. 

\item When only gravitational effects are concerned, the most complex evolutionary patterns in the ICM follow the crossing of shock waves (where the thermal and kinetic energy change significantly on a short timescale) and  turbulent motions (where the magnetic energy rapidly changes). Since the shock energy is more widely distributed than the turbulent energy in typical clusters at $z=0$, the profile of complexity is steeper in the case of magnetic energy than in the other two energy fields.

\item The magnetic energy displays the most complex behavior across most of the simulated cluster volume. Both in radiative and non-radiative simulations, describing the evolution of the magnetic field requires $\sim 2$ more information (in bits) than to describe the evolution of the thermal or of the kinetic energy, due to the presence of small-scale dynamo amplification in turbulent flows.

\item Radiative cooling and AGN feedback add significant complexity to the evolution of the ICM at all epochs.  By using the block entropy statistics and the entropy gain, we identify the emergence of extra complexity at high redshift ($z \geq 1$), before the virialization process of the gas in-falling onto the forming cluster proceeds. While the global energy statistics of the ICM at lower redshift are similar in the two cases, the block entropy carries memory of the complexity associated to each different process, acting on different epochs.

\item The efficiency of prediction (Sec.\ref{subsubsec:eff}) of the simulated ICM is largest at scales of $\sim 63-95$ kpc, consistent with the fact that turbulent and magnetic eddies in the simulated ICM have typical scales of this order. Future work will investigate in detail the dependence of this result against the data sampling strategy and the variety across the distribution of cluster dynamical states.

\end{itemize}

In conclusion, this first exploratory work shows that Information Theory has
the enormous potential of highlighting where, when and how cosmic structures become complex, and which physical ingredients
are more responsible for this. 
In particular, through this powerful tool  the {\it complexity} of a simulated Universe can be regarded as a well defined and measurable
field, which can be even visualized and followed in time.
While cosmological simulations will continue to increase the number of interconnected physical ingredients that can be
simulated at the same time, the development of the techniques suggested here will represent a 
powerful way for a deeper understanding of how structures emerge at all scales in the simulated Universe, and possibly in the real one.

\section{acknowledgments}

Computations described in this work were performed using the {\enzo} code (http://enzo-project.org), which is the product of a collaborative effort of scientists at many universities and national laboratories. 
I acknowledge the  usage of computational resources on the JURECA cluster at the at the Juelich Supercomputing Centre (JSC), under projects no. 7006 and 9016. I acknowledge personal support from the grant VA 876/3-1 and from the grant FOR1254  from the Deutsche Forschungsgemeinschaft (DFG). I acknowledge funding from the European Union's Horizon 2020 research and innovation programme under the Marie-Sklodowska-Curie gran agreement No 664931.
I gratefully acknowledge useful feedback from the anonymous referee of this paper, which resulted into a better presentation of these results.
I thank Annalisa and Leonardo for giving me the necessary ``free'' time that brought me the inspiration for this work.

\bibliographystyle{mnras}
\bibliography{info,franco}

\appendix

\section{Variations of the algorithms}

\begin{figure}
\includegraphics[width=0.459\textwidth]{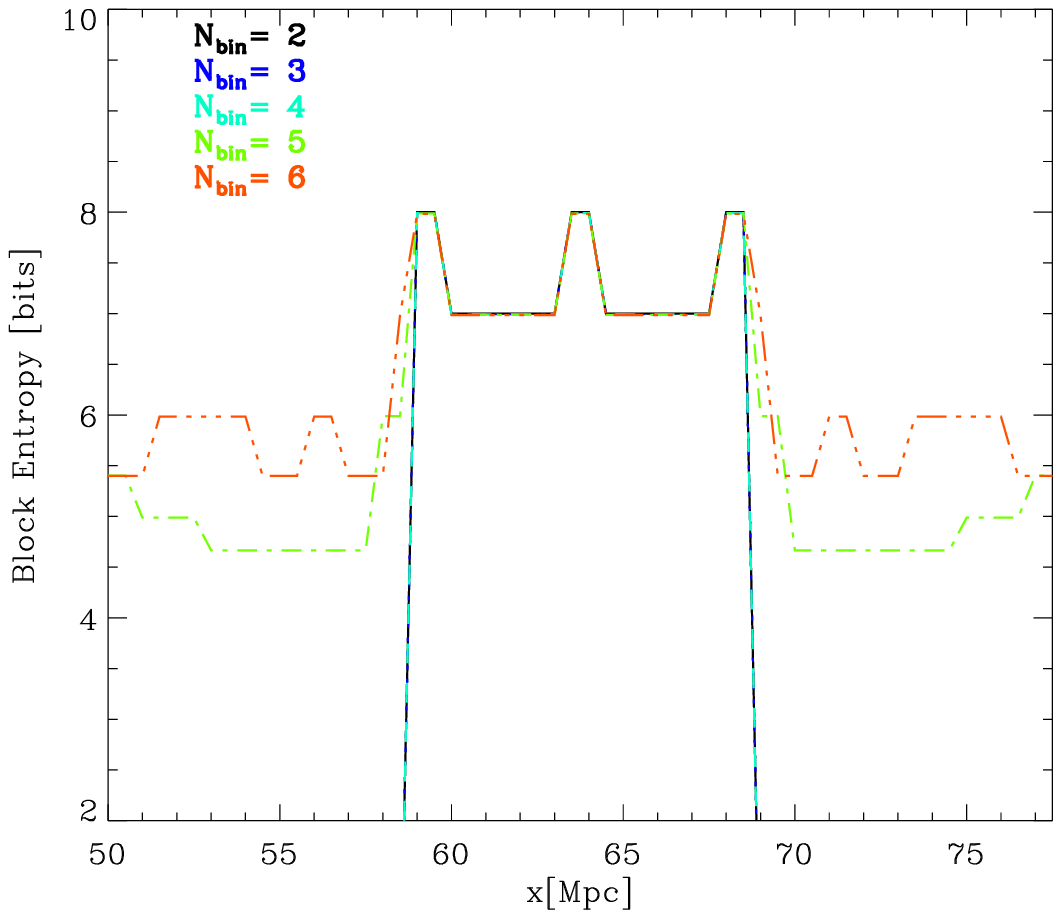}
\caption{Evolution of the block entropy in the Zeldovich collapse test, for different energy bins to compute $H(L)$.} 
 \label{fig:zeld3}
\end{figure}

The choice of the number of logarithmic energy bins and of the temporal and volume sampling of the datastream to compute statistical complexity and the block entropy are free parameters in our method. Here we give additional tests were we vary the fiducial parameters used in the main text, in order to show how the main conclusions of the work are overall unaffected by different choices. \\

Figure \ref{fig:zeld3} shows the profile of block entropy for the thermal energy in the Zeldovich pancake (Sec.\ref{zeld}) using 
 different number of bins for the coarse graining of $log_{\rm 10}(E_{\rm T})$.  The complexity within the pancake is a very robust measure against different choices for the number of bins, and in particular the 3 peaks mentioned above are independent of this. On the other hand, the degree of complexity in the rarefied gas outside of the pancake increases if the 
energy levels get $N_{\rm bin} \geq 5$, because of the adiabatic decrease in $E_T$. As also noted in Sec.\ref{zeld}, a larger number of energy bin can also highlight spurious fluctuations of numerical origin, which mostly occur in (energetically unimportant) regions undergoing rarefaction, and/or are associated  to the mesh de-refining procedure (Sec.\ref{subsec:non-rad}). We conclude that in general the specific choice of the energy binning strategy do not affect the most prominent  maxima of block entropy identified by our method. 

\bigskip

In Figure \ref{fig:block1_appendix} we study the block entropy for the 3 energy fields, as well as the ratio between the block entropy of the different energy fields and the total block entropy  for our non-radiative run, for 4 different spatial domains and/or choices of the time sampling of the datastream: a) 10 energy bins, 193,600 cells (fiducial run used in the main paper); b) 10 energy bins, 65,536 cells ; c) 10 energy bins, 48,400 cells, temporal distribution sampled every 2 root grid timesteps (the run was stopped at $\approx 9 \rm ~Gyr$ for memory requirements); d) 10 energy bins, 4096 cells. 

Although the global trend of the block entropy for the various energies is similar for these parameters variations, significant differences are present in the absolute value of block entropy at a given time. For example,  the run with the most refined time sampling (run C) gives the fastest increase in block entropy, owing to the larger variations in symbol statistics measured already at early times. While the absolute level of block entropy can vary by a factor $\sim 2-3$ at any specific epoch depending on the sampling strategy, we
find that the ratio of block entropy between the different fields gives a more robust view on the relative complexity of processes in
different epoch (Fig. \ref{fig:block_appendix}), as discussed in the main paper (Sec.\ref{subsec:rad}). 
In particular, the ratios converge to the values for $t \geq 6 \rm Gyr$, i.e. after the cluster fully formed, while show some small $\pm 1 \rm ~Gyr$ shift at earlier epochs, depending on how well is the forming cluster volume sampled by the different choices of volume.\\
 
Overall, we conclude that while the absolute level of the block entropy of each field obviously depends on the energy binning and on the adopted sampling, the ratios of complexity in the investigated energy fields is rather insensitive to this and the findings of the main paper are therefore robust. 

Finally, we estimated the error associated to the block entropy in our runs, due to the sampling variance inside the volume. To this
end, we followed the block entropy for 6 independent 2-dimensional slices through the cluster centre in the radiative run and using the 
interpolation to the resolution of $63$ kpc, which gives the best efficiency of prediction (Sec.\ref{eff_icm}).  The results are given in Fig.\ref{fig:errors}: while there are small variations due to the different sampling from slice to slice, the general trends are robust and are little
affected by the sample variance. More importantly,  the difference in complexity between the fields are in general much larger than the
sample variance, and the trends discussed in the main paper are robust against this.

\begin{figure}
\includegraphics[width=0.459\textwidth]{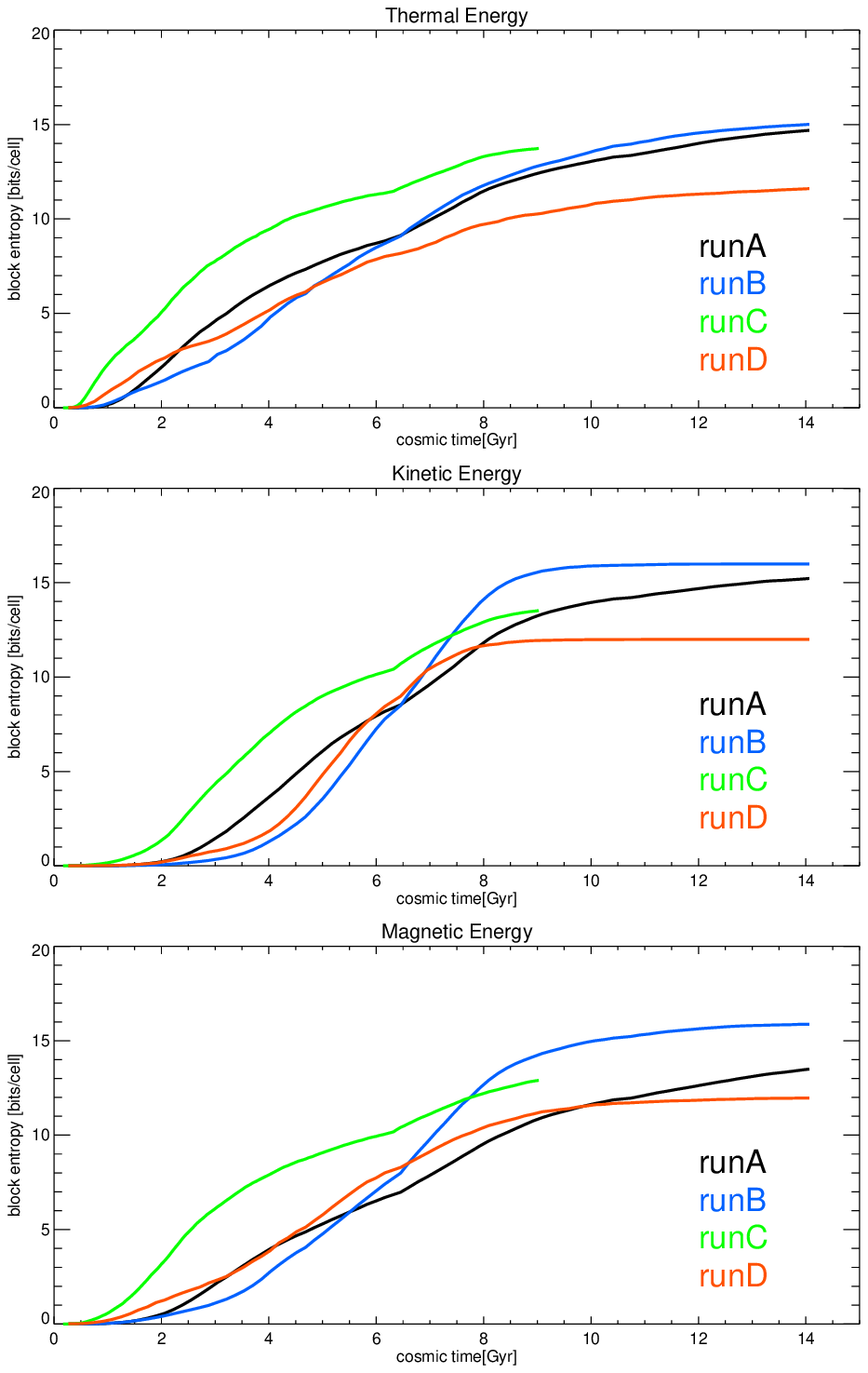}
\caption{Evolution of the block entropy of all simulated energy fields, for 4 different choices of energy binning and spatial domain (see text for details).}
 \label{fig:block1_appendix}
\end{figure}

\begin{figure}
\includegraphics[width=0.459\textwidth]{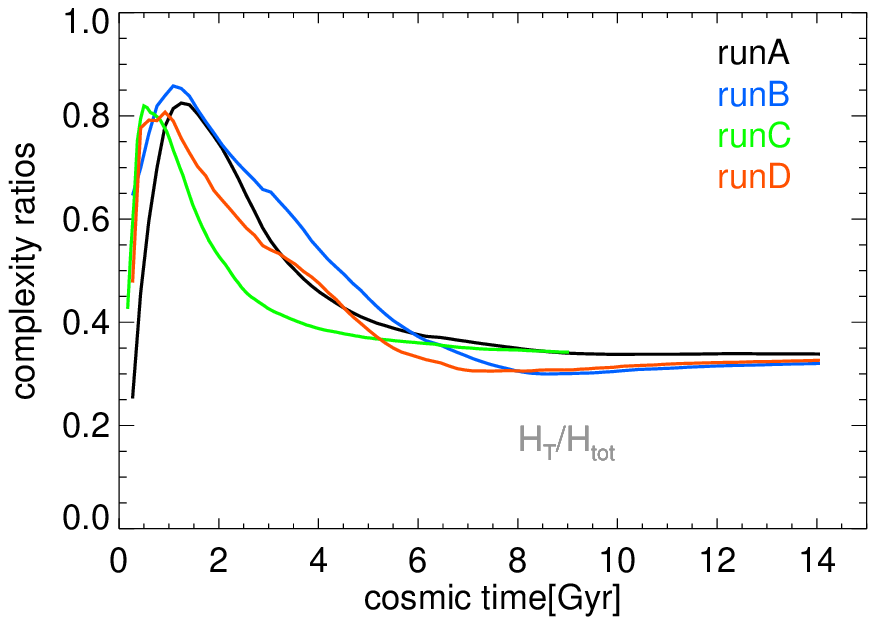}
\includegraphics[width=0.459\textwidth]{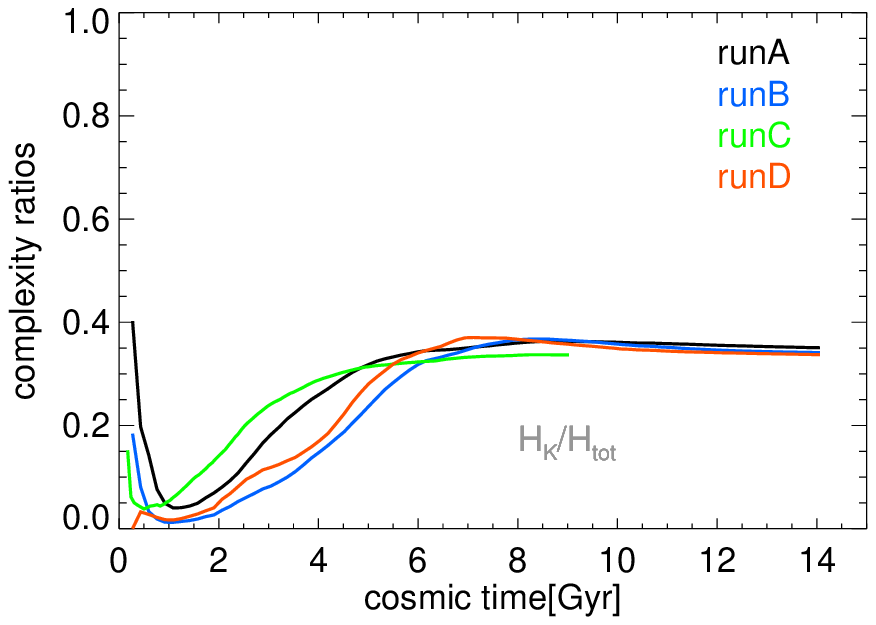}
\includegraphics[width=0.459\textwidth]{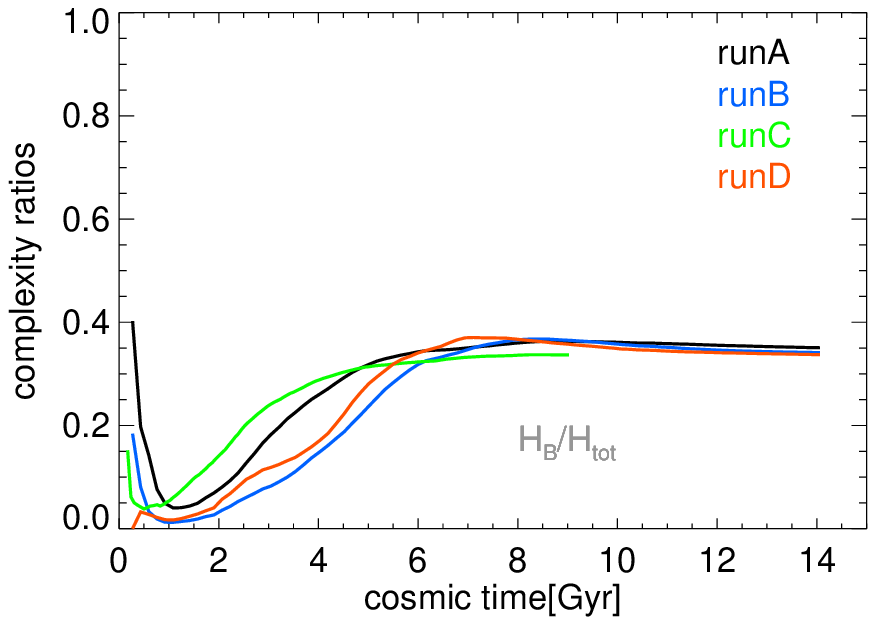}
\caption{Evolution of the ratio between the block entropy of each energy field and of the total block entropy, for the same energy
fields of Fig.\ref{fig:block1_appendix}.}
 \label{fig:block_appendix}
\end{figure}

\begin{figure}
\includegraphics[width=0.459\textwidth]{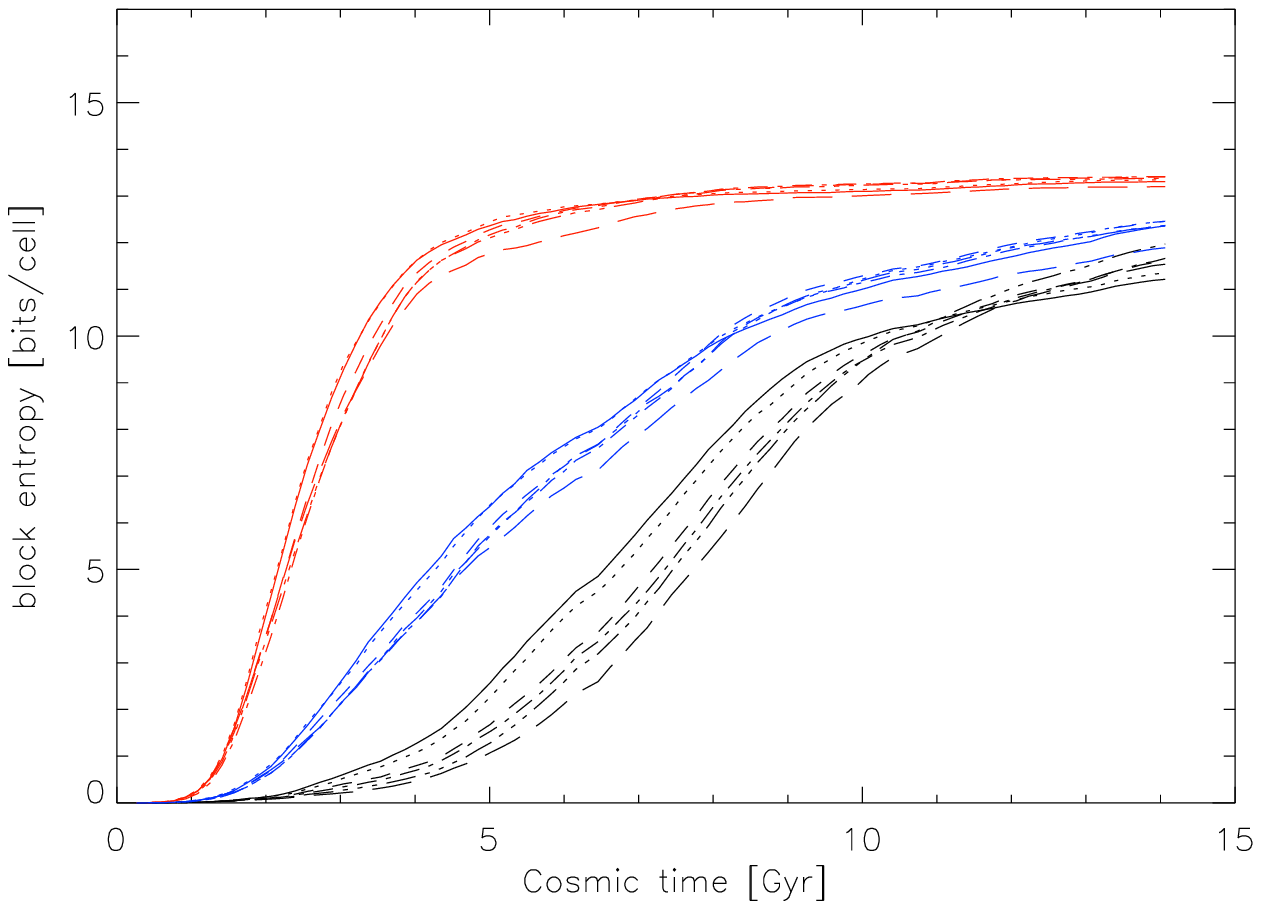}
\caption{Evolution of the block entropy of all simulated energy fields, for 6 different 2-dimensional slices through the center of the simulated cluster, with data binned to the spatial resolution of $63$ kpc.}
 \label{fig:errors}
\end{figure}
\end{document}